\documentclass{emulateapj}

\usepackage{amsmath}

\slugcomment{Accepted for publication in the Astrophysical
 Journal Supplement Series.}

\newcommand{\etal}{et~al.}
\newcommand{\eg}{e.g., }
\newcommand{\ie}{i.e., }

\newcommand{\Msun}{M_{\odot}}
\newcommand{\Rsun}{R_{\odot}}
\newcommand{\Lsun}{L_{\odot}}

\newcommand{\Ms}{M_{\rm preSN}}
\newcommand{\Rs}{R_{\rm preSN}}
\newcommand{\Rph}{R_{\rm ph}}
\newcommand{\Ls}{L_{\rm preSN}}
\newcommand{\Lp}{L_{\rm peak}}
\newcommand{\Tp}{T_{\rm c,peak}}
\newcommand{\Tmm}{T_{\rm MM99}}
\newcommand{\tmm}{t_{\rm MM99}}
\newcommand{\Lmm}{L_{\rm MM99}}
\newcommand{\Emm}{E_{\rm MM99}}
\newcommand{\Erad}{E_{\rm rad,1mag}}
\newcommand{\mlim}{m_{x,{\rm lim}}}
\newcommand{\mglim}{m_{g',{\rm lim}}}
\newcommand{\zlim}{z_{x,f}}

\newcommand{\thalf}{t_{\rm 1mag}}

\newcommand{\ergs}{erg~s$^{-1}$}

\newcommand{\Nifs}{$^{56}$Ni}
\newcommand{\Mms}{M_{\rm MS}}
\newcommand{\Mni}{M{\rm (^{56}Ni)}}
\newcommand{\Mej}{M_{\rm ej}}

\newcommand{\SN}{SNLS-04D2dc}
\newcommand{\Ebvh}{E(B-V)_{{\rm host}}}

\newcommand{\tobs}{t_{{\rm obs}}}
\newcommand{\tbobs}{t^{color}_{{\rm obs}}}
\newcommand{\tgrobs}{t^{g'-r'}_{{\rm obs}}}
\def\gsim{\mathrel{\rlap{\lower 4pt \hbox{\hskip 1pt $\sim$}}\raise 1pt
\hbox {$>$}}}
\def\lsim{\mathrel{\rlap{\lower 4pt \hbox{\hskip 1pt $\sim$}}\raise 1pt
\hbox {$<$}}}

\begin{document}

\title{Shock Breakout in Type II Plateau Supernovae:
Prospects for High Redshift Supernova Surveys}

\author{
 N.~Tominaga\altaffilmark{1,2},
 T.~Morokuma\altaffilmark{3,4,8},
 S.~I.~Blinnikov\altaffilmark{5,2},
 P.~Baklanov\altaffilmark{5},
 E.~I.~Sorokina\altaffilmark{6},
 K.~Nomoto\altaffilmark{2,7}
 }

\altaffiltext{1}{Department of Physics, Faculty of Science and
Engineering, Konan University, 8-9-1 Okamoto,
Kobe, Hyogo 658-8501, Japan; tominaga@konan-u.ac.jp}
\altaffiltext{2}{Institute for the Physics and Mathematics of the
Universe, University of Tokyo, 5-1-5 Kashiwanoha, Kashiwa, Chiba
277-8583, Japan}
\altaffiltext{3}{Institute of Astronomy, University of Tokyo, Mitaka, Tokyo 181-0015, Japan; tmorokuma@ioa.s.u-tokyo.ac.jp}
\altaffiltext{4}{Optical and Infrared Astronomy Division, National
Astronomical Observatory, 2-21-1 Osawa, Mitaka, Tokyo 181-8588, Japan}
\altaffiltext{5}{Institute for Theoretical and  Experimental Physics (ITEP),
Moscow 117218, Russia; sergei.blinnikov@itep.ru, baklanovp@gmail.com}
\altaffiltext{6}{Sternberg
Astronomical Institute, Moscow 119992, Russia; sorokina@sai.msu.su}
\altaffiltext{7}{Department of Astronomy, School of Science,
University of Tokyo, Bunkyo-ku, Tokyo 113-0033, Japan;
nomoto@astron.s.u-tokyo.ac.jp}
\altaffiltext{8}{Research Fellow of the Japan Society for the Promotion of Science}

\begin{abstract}

 Shock breakout is the brightest radiative phenomenon in a
 supernova (SN) but is difficult to be observed
 owing to the short duration and X-ray/ultraviolet (UV)-peaked
 spectra. After the first observation from the rising
 phase reported in 2008, its
 observability at high redshift is attracting enormous attention. 
 We perform multigroup radiation hydrodynamics calculations of
 explosions for evolutionary presupernova models with various
 main-sequence masses $\Mms$, metallicities $Z$, and explosion energies
 $E$. We present multicolor light curves of shock breakout in Type II
 plateau SNe, being the most frequent core-collapse SNe, and predict
 apparent multicolor light curves of shock breakout at
 various redshifts $z$. We derive the observable SN rate and reachable
 redshift as functions of filter $x$ and limiting magnitude
 $\mlim$ by taking into account an initial mass function, cosmic
 star formation history, intergalactic absorption, and host galaxy
 extinction. We propose a realistic
 survey strategy optimized for shock breakout. For
 example, the $g'$-band observable SN
 rate for $\mglim=27.5$~mag is $3.3$~SNe~degree$^{-2}$~day$^{-1}$ and a half
 of them locates at $z\geq1.2$. It is clear that the shock breakout
 is a beneficial clue to probe high-$z$ core-collapse SNe. We also establish ways to identify
 shock breakout and constrain SN properties from the observations
 of shock breakout, brightness, time scale, and color. We 
 emphasize that the multicolor observations in blue
 optical bands with $\sim$~hour intervals, preferably over $\geq2$ continuous nights,
 are essential to efficiently detect, identify, and interpret shock breakout.
\end{abstract}


\section{INTRODUCTION}
\label{sec:intro}

In contrast to Type Ia supernovae (SNe), systematic observational studies of
core-collapse supernovae (CCSNe) have been restricted at a redshift
$z\lsim1$ (\eg \citealt{poz07}). This is because CCSNe
are typically fainter than Type Ia SNe \citep[\eg][]{ric02},
except for rare energetic supernovae (hypernovae, \eg SN~1998bw,
\citealt{gal98}) or recently-found extremely bright SNe (\eg
\citealt{qui09}).

Recent improvements of telescopes/instruments and well-organized survey strategies make it
possible to detect unusual core-collapse events at high redshift: for
example, bright Type IIn SNe (SNe IIn) at $z>2$ \citep{coo09} and gamma-ray
bursts (GRBs) up to $z\sim8.2$ \citep{sal09,tan09}. With the use of
the high-$z$ accessibility of these events, a star formation history
(SFH) and an initial mass function (IMF) at high redshift are intensively studied (\eg 
\citealt{kis09,wan09,coo09}). However, special conditions are required to
realize such events, \ie a dense circumstellar matter for an SN IIn (\eg
\citealt{chu04}) and a fast-rotating progenitor for a GRB (\eg
\citealt{woo93}). Thus they can not be the main constituents of CCSNe
and the SFH and IMF estimated with them could
involve large biases. Hence, ways to directly detect normal CCSNe
at high redshift are required to investigate the nature of majority of
CCSNe and estimate the SFH and IMF with small biases.

The bolometrically-brightest phenomenon in the SN with a shockwave is shock
breakout. In a CCSN explosion, an outward shockwave forms around a
central remnant by depositing released gravitational energy. The
shockwave propagates through a stellar envelope to heat it up and
accelerates its expansion. Since the star is
optically thick, the shockwave cannot be electromagnetically
observed until its emergence from a stellar surface. When the
shockwave approaches the stellar surface at a distance with an optical depth $\lsim10$,
radiation from the shock front starts to leak out and a hot fire ball
suddenly appears to emit a bright soft X-ray and ultraviolet (UV) flash
with a quasi-blackbody spectrum ($T>10^5$~K). This phenomenon is shock breakout
which has been theoretically predicted by, \eg \cite{kle78}. 
Its duration strongly depends on the presupernova radius. Brightness
rises in several seconds to several hours and declines in several ten
seconds to several days.
Theoretical studies have suggested that the peak bolometric luminosity exceeds 
$10^{44}$~erg~s$^{-1}$ (\eg \citealt{bli00}) and that shock
breakout is observable even if it takes place at $z\gsim1$ (\eg
\citealt{chu00}). However, the short duration and soft X-ray/UV-peaked
spectra had made detection of shock breakout difficult for a long while, except
for fortunate detection of its tail in nearby SNe in $U$ band
(\eg SN~1987A, \citealt{cat87}; SN~1993J, \citealt{ric94}; SN~1999ex, \citealt{str02}).

First detection of shock breakout in the rising phase was obtained
serendipitously in 2004--2008 and reported in 2008:
Type Ib SN~2008D in NGC~2770 (distance $d=27$ Mpc,
\citealt{sod08,maz08,mod09,mal09}), Type II plateau SN (SN II-P)
SNLS-04D2dc (redshift $z=0.185$, \citealt{sch08,gez08}), and SN II-P SNLS-06D1jd
($z=0.324$, \citealt{gez08}). 

For SN~2008D, an X-ray LC from breakout and optical LCs after a tail
were observed (\eg \citealt{sod08,mod09}). The X-ray LC rises in
$\sim60$~sec and declines in $\sim130$~sec but it
is not clear whether the X-ray spectral energy distribution (SED) is
thermal or nonthermal. For the SNLS SNe~II-P, UV LCs of shock breakout
and optical LCs of plateau were
observed (\eg \citealt{sch08,gez08}). The UV flash of SNLS-04D2dc rises
and declines in several hours and the subsequent UV LC shows
rebrightening in several days. However, the UV data have too low
signal-to-noise ratios to obtain SED.\footnote{The UV
observations of SNLS-06D1jd have too poor signal-to-noise ratios to
extract the characteristics of shock breakout.} 

The observational papers and subsequent
papers present theoretical models for shock breakout;
analytic models for SN~2008D \citep[\eg][]{sod08,che08,mod09}\footnote{The subsequent LCs
and spectral evolution of SN~2008D are reproduced by the radiative transfer
calculation with homologous hydrodynamics evolution \citep[\eg][]{tanaka09b,tanaka09}.} and
hydrodynamics calculation with two-temperature radiative diffusion
\citep{sch08} and one-temperature radiation
hydrodynamics calculation coupled with non-local thermodynamic
equilibrium (non-LTE) spectral calculation for \SN\ \citep{gez08}. 
\cite{tom09b} have performed a multigroup radiation hydrodynamical
calculation for the first time and
successfully constructed a self-consistent radiation hydrodynamical model
for SNLS-04D2dc. They have presented that an SN explosion of a $20\Msun$ star
with an explosion energy $1.2\times10^{51}$~erg reproduces well
the UV-optical LCs of shock breakout and
plateau. Also, the successful model 
demonstrates that an SN similar to \SN\ is detectable at $z=1$ with
8m-class optical telescopes.

The UV-bright shock breakout is followed by a plateau phase.
It appears as an SN~II-P that is most frequent among core-collapse SNe (\eg
\citealt{man08,arc10,li10b,smi10}). In contrast to an SN~IIn and a GRB,
the formation of a plateau phase does 
not require any specified conditions other
than the presence of a thick H envelope. Therefore, the SFH estimated with shock
breakout in SNe~II-P should have smaller uncertainties and biases than those with SNe IIn or
GRBs. Moreover, observational properties of shock breakout, the
brightness, color, and time scale, depend on properties of the SN and its
progenitor, the explosion energy $E$, presupernova radius $\Rs$, and
ejecta mass $\Mej$ (\eg \citealt{mat99}). Thus, it is possible to
derive detailed properties of SN explosions from the observations of shock
breakout and constrain an IMF precisely.

\begin{figure}[t]
\plotone{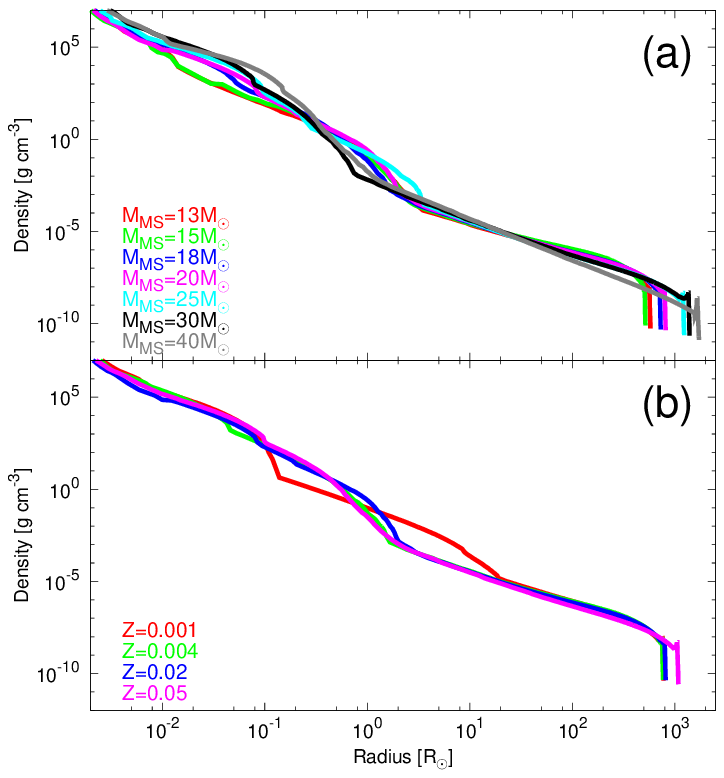} 
\caption{Presupernova density structures of (a) $Z=0.02$ stars with
 $\Mms=13\Msun$ ({\it red}), $\Mms=15\Msun$ ({\it green}),
 $\Mms=18\Msun$ ({\it blue}), $\Mms=20\Msun$ ({\it magenta}),
 $\Mms=25\Msun$ ({\it cyan}), $\Mms=30\Msun$ ({\it black}), and
 $\Mms=40\Msun$ ({\it gray}), and (b) $20\Msun$ models with 
 $Z=0.001$ ({\it red}), $Z=0.004$ ({\it green}), 
 $Z=0.02$ ({\it blue}), and $Z=0.05$ ({\it magenta}).
}
\label{fig:preSN}
\end{figure}

\begin{deluxetable}{c|cccc}[t]
 \tabletypesize{\scriptsize}
 \tablecaption{Progenitor models. \label{tab:preSN}}
 \tablewidth{0pt}
 \tablehead{
   \colhead{$\Mms$}
 & \colhead{$Z$}
 & \colhead{$\Ms$}
 & \colhead{$\Rs$}
 & \colhead{$\Ls$}\\
   \colhead{[$\Msun$]}
 & \colhead{}
 & \colhead{[$\Msun$]}
 & \colhead{[$\Rsun$]}
 & \colhead{[$10^4\Lsun$]}
 }
\startdata
 13& 0.02  & 12.7 & 564  & 5.57 \\
 15& 0.02  & 14.1 & 507  & 4.69 \\
 18& 0.02  & 16.7 & 713  & 8.81 \\
 20& 0.02  & 18.4 & 795  & 11.1 \\
 25& 0.02  & 21.7 & 1200 & 23.9 \\
 30& 0.02  & 25.0 & 1360 & 31.4 \\
 40& 0.02  & 21.7 & 1660 & 53.8 \\ 
 20& 0.001 & 19.7 & 756  & 18.6  \\
 20& 0.004 & 19.5 & 751  & 13.4 \\
 20& 0.05  & 17.3 & 1050 & 15.7 
\enddata
\end{deluxetable}

\begin{figure*}[t]
\epsscale{.8}
\plotone{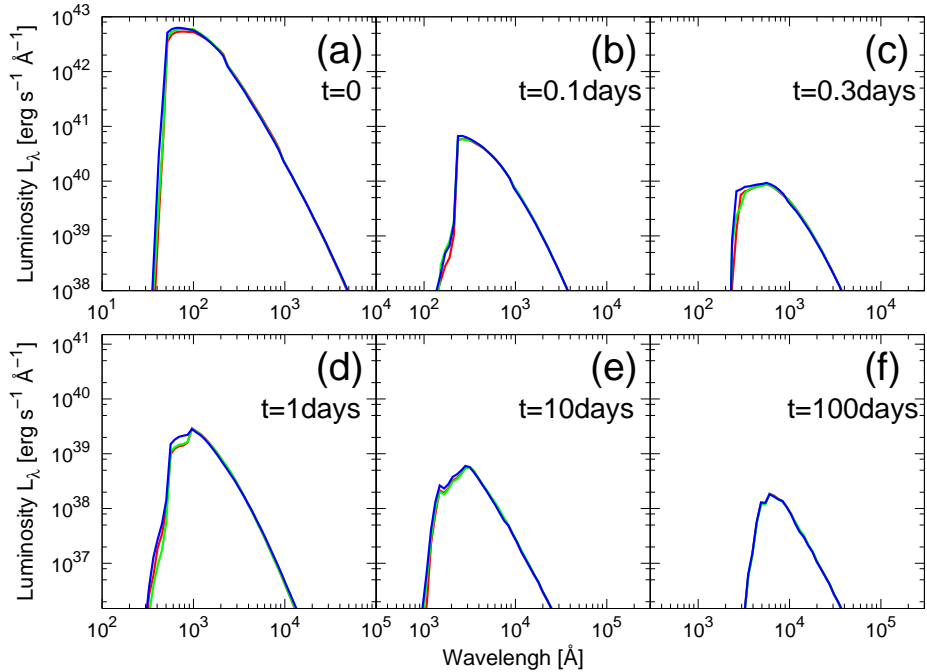} 
\caption{Spectral energy distributions at (a) $t=0$, (b) $t=0.1$~days,
 (c) $t=0.3$~days, (d) $t=1$~days, (e) $t=10$~days, and (f) $t=100$~days
 for different opacity prescriptions ({\it red}: original, {\it green}:
 including excited levels in bound-free absorption, and {\it blue}
 including inner-shell photo-ionization).
}
\label{fig:opacity}
\end{figure*}

Shock breakout is coming under the spotlight to probe high-$z$
CCSNe but the observable properties are poorly understood. Therefore, in
order to execute a shock breakout survey effectively, it is required
to provide theoretical predictions for observable quantities and
propose strategies of observations and analysis to detect, identify, and interpret them.
Hence, we perform multigroup radiation hydrodynamics calculations of shock
breakout in SNe~II-P with various $\Mms$, $Z$, and $E$ with a multigroup
radiation hydrodynamics code {\sc stella} \citep{bli98,bli00,bli06}
and present theoretical predictions of apparent multicolor LCs 
at various redshifts. Based on the theoretical models,
we estimate the number of detection and reachable redshift, clarify
requirements on survey strategies, 
and develop ways to identify shock breakout and to derive the SN
properties from the observational quantities of shock breakout.

In \S~\ref{sec:model}, the applied models and methods are briefly described. In
\S~\ref{sec:result}, results are shown. We present
the multicolor LCs of shock breakout (\S~\ref{sec:LC}) and 
predictions of apparent multicolor LCs of shock breakout (\S~\ref{sec:highz}).
In \S~\ref{sec:future} we offer future prospects on shock breakout
surveys: an expected number of detection and reachable redshift (\S~\ref{sec:number}),
dependencies on extinction and SFH (\S~\ref{sec:uncertain}), requirements
on survey strategies (\S~\ref{sec:realobs}), ways to identify
shock breakout (\S~\ref{sec:ID}), and ways to constrain SN properties
(\S~\ref{sec:SNprop}). In \S~\ref{sec:discuss}, the conclusion and
discussion are presented.

\section{Models \& Methods}
\label{sec:model}

\subsection{Progenitor model \& explosive nucleosynthesis}
\label{sec:preSN}

We adopt progenitor models with various $\Mms$
($=13,~15,~18,~20,~25,~30,~{\rm and}~40\Msun$) and $Z$ 
($=0.001,~0.004,~{\rm and}~0.02$) which are taken from
\cite{ume05a}.
The stellar evolution calculations include a mass loss depending on
metallicity $Z$ which is assumed to be proportional to $Z^{0.5}$
\citep{kud00a}. Since shock breakout arises at a thin surface
layer with an optical depth $\tau\lsim10$, we adopt the stellar surface
sufficiently outside which is as shallow as $\tau=0.001$.\footnote{In order to
justify the shallowness of the outer boundary, we perform a radiation
hydrodynamic calculation for an envelope model down to $\tau=10^{-8}$
and confirm the consistency with the envelope model down to
$\tau=0.001$. On the other hand, we find that an envelope model down to
$\tau=0.01$ shows a bluer SED  than that down to $\tau=0.001$. This
indicates that $\tau=0.01$ is not shallow enough to treat SEDs of shock
breakout.}
The density structures of the progenitor models are shown
in Figures~\ref{fig:preSN}a and \ref{fig:preSN}b. 
The properties of progenitor models, $\Mms$, $Z$, presupernova mass $\Ms$,
presupernova radius $\Rs$, and presupernova luminosity $\Ls$, are summarized in
Table~\ref{tab:preSN}.\footnote{We note that the progenitor
structures, \ie the relations among $\Mms$, $Z$, $\Ms$, $\Rs$, and $\Ls$,
depend on the treatment of physics, \eg rotation, pulsation,
mass loss, mixing length, and overshooting (\eg \citealt{lim06,yoo10}).} 
$\Rs$, being the most important ingredient for
shock breakout, increases roughly monotonically with $Z$ and
$\Mms$, except for the model with $\Mms=15\Msun$. The progenitor models
are red supergiants with $\Rs=500-1700\Rsun$ producing SNe~II-P.
We note that the density inversion
at the outermost layer corresponds to a super-adiabatic layer, where the
temperature gradient is steeper than the adiabatic case.

We calculate explosive nucleosynthesis adopting the same method as
described in \cite{tom07b}; the explosion is initiated as a thermal
bomb, hydrodynamics is calculated including nuclear energy generation
with the $\alpha$-network, and a nucleosynthesis calculation is performed
as a postprocessing. Since explosive nucleosynthesis ceases well before the
shock emergence from the stellar surface, we perform a radiation
hydrodynamics calculation of an explosion for a model with the abundance distribution
after explosive nucleosynthesis and the hydrodynamical structure of the
progenitor model. 

\subsection{Radiation hydrodynamics}

\begin{figure*}[t]
\plotone{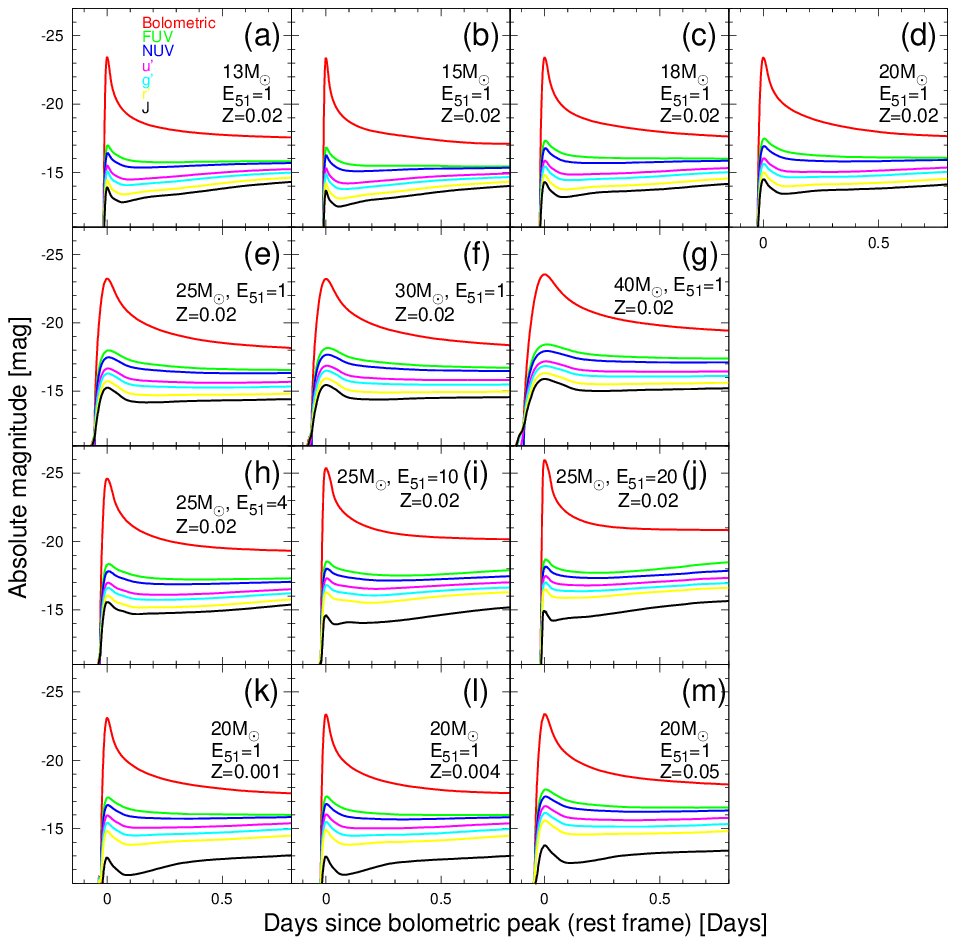} 
\caption{Absolute bolometric ({\it red}) and multicolor LCs
({\it green}: FUV, {\it blue}: NUV, {\it magenta}:
 $u'$, {\it cyan}: $g'$, {\it yellow}: $r'$, and {\it black}: $J$) of
the models with 
(a) $\Mms=13\Msun$, $Z=0.02$, $E_{51}=1$,
(b) $\Mms=15\Msun$, $Z=0.02$, $E_{51}=1$,
(c) $\Mms=18\Msun$, $Z=0.02$, $E_{51}=1$,
(d) $\Mms=20\Msun$, $Z=0.02$, $E_{51}=1$,
(e) $\Mms=25\Msun$, $Z=0.02$, $E_{51}=1$,
(f) $\Mms=30\Msun$, $Z=0.02$, $E_{51}=1$,
(g) $\Mms=40\Msun$, $Z=0.02$, $E_{51}=1$,
(h) $\Mms=25\Msun$, $Z=0.02$, $E_{51}=4$,
(i) $\Mms=25\Msun$, $Z=0.02$, $E_{51}=10$,
(j) $\Mms=25\Msun$, $Z=0.02$, $E_{51}=20$,
(k) $\Mms=20\Msun$, $Z=0.001$, $E_{51}=1$,
(l) $\Mms=20\Msun$, $Z=0.004$, $E_{51}=1$, and
(m) $\Mms=20\Msun$, $Z=0.05$, $E_{51}=1$.
}
\label{fig:AbsLC}
\end{figure*}

We use the multigroup radiation hydrodynamics code {\sc stella}
\citep{bli98,bli00,bli06}. The detail of {\sc stella} is
described in archival literatures \citep{bli98,bli00,bli06} and its reliability
has been carefully verified by comparisons with analytic solutions
\citep{mat99,rab10}, other numerical codes \citep{bli98,bli03}, and
multicolor SN observations \citep{bli98,bli00,bli06,chu04}. Here, we briefly
describe the assumptions and procedures applied in {\sc stella} and the
setup adopted in this paper. 

{\sc stella} solves the time-dependent
equations implicitly for the angular moments of intensity averaged over
fixed frequency bands and computes variable Eddington
factors that fully take into account scattering and
redshifts for each frequency group in each mass zone. The $\gamma$-ray
transfer is calculated using a one-group approximation for the nonlocal
deposition of the energy of radioactive nuclei; we follow Swartz
\etal\ (1995; see also \citealt{jef98}) and use a purely
absorptive opacity for $\gamma$-ray. It is worthy to note that the
$\gamma$-ray transfer does not influence the results in this paper
because of no contribution to shock breakout from
the radioactive decays. In the equation of state, LTE ionizations and recombinations
are taken into account. The effect of line opacity is
treated as an expansion opacity according to the prescription
of Eastman \& Pinto (1993; see also \citealt{bli98}). 

We adopt 100 frequency bins dividing
logarithmically from $\nu=6\times10^{13}$~Hz
($\lambda=5\times10^4$~\AA) to $3\times10^{18}$~Hz (1~\AA); such a
number of frequency bins are enough to solve non-equilibrium continuum
radiation and treat any SEDs accurately. We emphasize that there
is no need to ascribe any temperature to the radiation.
The coupling of multigroup radiation transfer with hydrodynamics enables
us to obtain the color temperature in a self-consistent calculation, \ie
a luminosity-weighted blackbody fitting of SED. 

\begin{figure*}[t]
\epsscale{.8}
\plotone{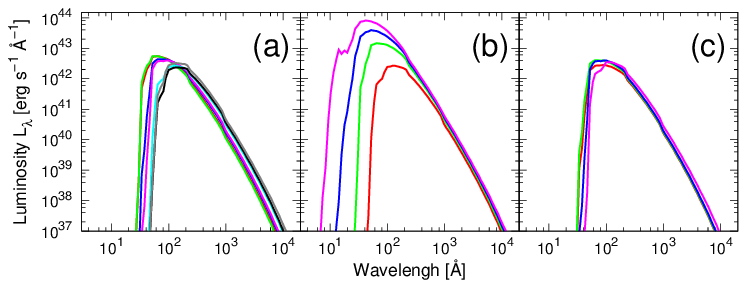} 
\caption{Spectral energy distributions at $t=0$ of the models with (a)  
 $\Mms=13\Msun$, $Z=0.02$, $E_{51}=1$ ({\it red}),
 $\Mms=15\Msun$, $Z=0.02$, $E_{51}=1$ ({\it green}),
 $\Mms=18\Msun$, $Z=0.02$, $E_{51}=1$ ({\it blue}),
 $\Mms=20\Msun$, $Z=0.02$, $E_{51}=1$ ({\it magenta}),
 $\Mms=25\Msun$, $Z=0.02$, $E_{51}=1$ ({\it cyan}),
 $\Mms=30\Msun$, $Z=0.02$, $E_{51}=1$ ({\it black}), and
 $\Mms=40\Msun$, $Z=0.02$, $E_{51}=1$ ({\it gray}), 
 (b) $\Mms=25\Msun$, $Z=0.02$, $E_{51}=1$ ({\it red}),
 $\Mms=25\Msun$, $Z=0.02$, $E_{51}=4$ ({\it green}),
 $\Mms=25\Msun$, $Z=0.02$, $E_{51}=10$ ({\it blue}), and
 $\Mms=25\Msun$, $Z=0.02$, $E_{51}=20$ ({\it magenta}),
 (c) $\Mms=20\Msun$, $Z=0.001$, $E_{51}=1$ ({\it red}),
 $\Mms=20\Msun$, $Z=0.004$, $E_{51}=1$ ({\it green}),
 $\Mms=20\Msun$, $Z=0.02$, $E_{51}=1$ ({\it blue}), and
 $\Mms=20\Msun$, $Z=0.05$, $E_{51}=1$ ({\it magenta}).
}
\label{fig:color}
\end{figure*}

\begin{figure*}[t]
\epsscale{.8}
\plotone{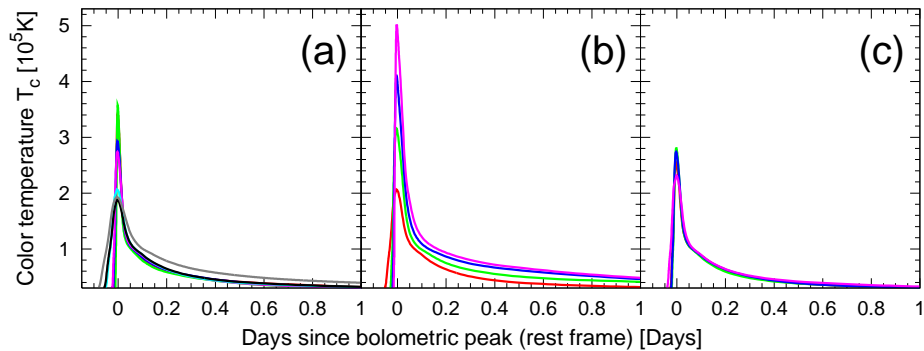} 
\caption{Color temperature evolution of the models. The colors and panels are the 
same as in Figure~\ref{fig:color}.
}
\label{fig:temp}
\end{figure*}

All previous computations with {\sc stella} employed the assumptions
used in the code {\sc eddington} \citep{eas93} for bound-free
transitions, in which all atoms and ions, except
for hydrogen, are in ground states.
Since new opacity tables for {\sc stella} will be released including
excited levels in bound-free absorption (E.~Sorokina in prep.) and
inner-shell photo-ionization (P.~Baklanov in prep.), we briefly examine
these effects on SEDs for a model with
$\Mms=20\Msun$, $Z=0.02$, and explosion energy 
$E_{51}=E/(10^{51}~{\rm erg})=1$.
While the inner-shell photo-ionization cross-sections are based on
formulae derived by \cite{ver93,ver96b} and \cite{ver95}\footnote{See 
\url{http://www.pa.uky.edu/\~ verner/photo.html} for details.} as in old
{\sc eddington} and {\sc stella} routines,
the case with excited levels treats all bound-free
transitions, also for ground-levels, by different fitting
formulae as in code {\sc
wmbasic}\footnote{\url{http://www.usm.uni-muenchen.de/people/adi/Programs/Programs.html}} (\citealt{pau87}, see also E.~Sorokina in prep.).

The SEDs
with three different opacity prescriptions are shown in
Figure~\ref{fig:opacity}. Although there are small differences
as $\leq0.3$~mag in several frequency bins, the differences are diluted
and diminished when
SEDs are convolved with broad-band filters.
This illustrates that {\sc stella} results are robust with respect
to different approximations for bound-free transitions at least for
shock breakout in an SN~II-P in which temperature is not extremely high
($T<10^6$~K). Therefore, in this paper we adopt a procedure from \cite{eas93}.
The opacity table includes $1.5\times10^5$ spectral lines from
\cite{kur95} and \cite{ver96a}. 

\section{Results}
\label{sec:result}

\subsection{Synthetic multigroup light curves}
\label{sec:LC}

We perform the multigroup radiation hydrodynamics calculations for
models with various $\Mms$, $E$, and $Z$. Here, for simplicity, we set
$\Mej$ to yield a canonical amount of \Nifs\ without
mixing to the envelope [the ejected \Nifs\ mass $\Mni=0.07\Msun$]. 
We note that the assumption does not
affect the results because the radioactive decays
do not contribute to shock breakout. In this paper, we adopt the
AB magnitude system and the following filters: far UV (FUV) and near UV (NUV)
bands for {\it GALEX} satellite \citep{mor05,mor07}, $u'$, $g'$,
$r'$, $i'$ and $z'$ bands \citep{fuk96}, $J$, $H$, and $K$ bands
\citep{tok02}, and F322W band for {\it The James Webb Space Telescope}
({\it JWST}, \citealt{gar06}). 

The multigroup radiation hydrodynamics calculation provides wavelength-
and time-dependent fluxes at an SN surface. Since lights radiated from
different parts of the SN surface arrive at an observer in a given
direction at different time (for details, see
\citealt{kle78,ims81,ens92,bli02,bli03}), a light
travel time correction\footnote{The light travel time effect in an aspherical explosion is
investigated in, \eg \cite{cou09} and \cite{suz10}.} and a limb darkening
correction in the Eddington approximation are applied.

\begin{deluxetable*}{cccc|cccc|cccc}[t]
 \tabletypesize{\scriptsize}
 \tablecaption{Properties of shock breakout. \label{tab:SB}}
 \tablewidth{0pt}
 \tablehead{
   \colhead{$\Mms$}
 & \colhead{$Z$}
 & \colhead{$E$}
 & \colhead{$\Mej$}
 & \colhead{$\Tp$}
 & \colhead{$\thalf$}
 & \colhead{$\Erad$}
 & \colhead{$\Lp$}
 & \colhead{$\Tp$}
 & \colhead{$\thalf$}
 & \colhead{$\Erad$}
 & \colhead{$\Lp$}\\
   \colhead{[$\Msun$]}
 & \colhead{}
 & \colhead{[$10^{51}$~erg]}
 & \colhead{[$\Msun$]}
 & \colhead{[$10^{5}$~K]}
 & \colhead{[$10^{-2}$~days]}
 & \colhead{[$10^{48} {\rm erg}$]}
 & \colhead{[$10^{44} {\rm erg~s^{-1}}$]}
 & \colhead{[$10^{5}$~K]}
 & \colhead{[$10^{-2}$~days]}
 & \colhead{[$10^{48} {\rm erg}$]}
 & \colhead{[$10^{44} {\rm erg~s^{-1}}$]}\\
   \colhead{}
 & \colhead{}
 & \colhead{}
 & \colhead{}
 & \multicolumn{4}{c}{without corrections}
 & \multicolumn{4}{c}{with corrections}
}

\startdata
 13& 0.02  & 1  & 11.2 & 3.56  & 0.651 & 0.691 & 10.3  & 3.35 & 1.09 & 0.815 & 6.99 \\
 15& 0.02  & 1  & 12.7 & 3.68  & 0.512 & 0.536 & 10.2  & 3.48 & 0.915& 0.643 & 6.64 \\
 18& 0.02  & 1  & 15.2 & 3.03  & 0.994 & 0.948 & 9.41  & 2.87 & 1.46 & 1.09  & 6.90 \\
 20& 0.02  & 1  & 16.8 & 2.80  & 1.19  & 1.11  & 9.11  & 2.70 & 1.68 & 1.26 & 6.91 \\
 25& 0.02  & 1  & 19.9 & 2.17  & 2.94  & 2.07  & 6.78  & 2.04 & 3.40 & 2.23 & 5.99 \\
 30& 0.02  & 1  & 23.0 & 1.99  & 3.51  & 2.39  & 6.52  & 1.87 & 3.97 & 2.54 & 5.86 \\
 40& 0.02  & 1  & 19.6 & 2.01  & 5.16  & 4.57  & 8.55  & 1.92 & 5.60 & 4.82 & 7.90 \\ 
 25& 0.02  & 4  & 19.8 & 3.23  & 1.27  & 4.20  & 32.5  & 3.08 & 2.28 & 5.05 & 21.1 \\
 25& 0.02  & 10 & 19.7 & 4.38  & 0.763 & 6.83  & 89.6  & 3.96 & 2.12 & 8.74 & 42.4 \\
 25& 0.02  & 20 & 19.6 & 5.21  & 0.511 & 10.0  & 196   & 4.88 & 2.06 & 13.5 & 72.4 \\
 20& 0.001 & 1  & 17.9 & 2.57  & 1.35  & 0.896 & 6.51  & 2.51 & 1.77 & 1.01 & 5.23 \\
 20& 0.004 & 1  & 17.8 & 2.91  & 1.09  & 0.986 & 8.80  & 2.76 & 1.54 & 1.12 & 6.64 \\
 20& 0.05  & 1  & 15.7 & 2.41  & 2.14  & 1.80  & 8.09  & 2.31 & 2.64 & 1.99 & 6.88
\enddata

\end{deluxetable*}

\begin{figure}[t]
\epsscale{1.}
\plotone{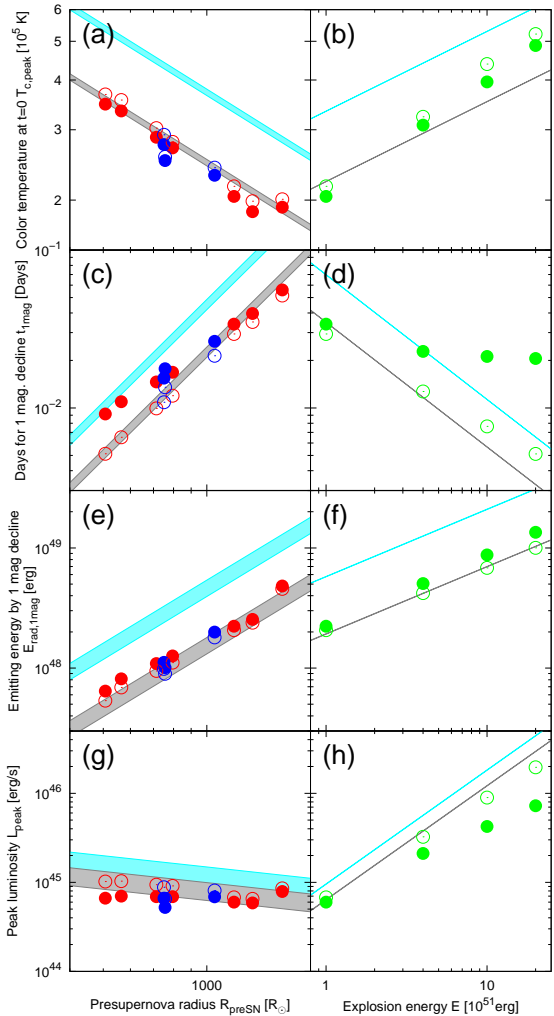}
\caption{Comparisons among the models with corrections ({\it filled circles}),
 the models without corrections ({\it open circles}), and semi-analytic solutions
 (original: {\it cyan shaded region} and reduced: {\it gray shaded
 region}). The color of symbols represents $Z=0.02$ models ({\it red}),
 $\Mms=20\Msun$ models with $Z=0.001,~0.004,~{\rm and}~0.05$ ({\it
 blue}), and $\Mms=25\Msun$ models with $E_{51}=1,~4,~10,~{\rm and}~20$
 ({\it green}). 
}
\label{fig:SB}
\end{figure}

Absolute LCs of the corrected models are shown in
Figures~\ref{fig:AbsLC}a-\ref{fig:AbsLC}m.
Figures~\ref{fig:AbsLC}a-\ref{fig:AbsLC}g,
\ref{fig:AbsLC}h-\ref{fig:AbsLC}j, and \ref{fig:AbsLC}k-\ref{fig:AbsLC}m
show the LCs of the models with different $\Mms$, different $E$, and
different $Z$, respectively. The models with larger $\Mms$
and higher $Z$ have larger $\Rs$, except for the model with
$\Mms=15\Msun$, thus having broader and slightly fainter peak. On the other
hand, the models with higher $E$ have brighter and narrower peak. While
the bolometric luminosities decline monotonically with time due to an
adiabatic cooling, there could be rebrightening in homochromatic LCs due
to the shift of peak wavelength with time.

The SEDs at the
bolometric peak ($t=0$, hereafter we set $t=0$ at the bolometric peak of
each model) are shown in Figures~\ref{fig:color}a-\ref{fig:color}c. The
SEDs at $t=0$ peak in UV ($\lambda\sim40-100$\AA) and have similar
spectral slopes at
$\lambda\gsim400$\AA, while the luminosities at $\lambda\gsim400$\AA\
are higher for larger $\Rs$ and slightly higher for higher $E$. Evolution of color
temperature is shown in Figures~\ref{fig:temp}a-\ref{fig:temp}c. The
color temperatures range from $T_{\rm c}\sim2\times10^5$ to 
$T_{\rm c}\sim5\times10^5$~K
at $t=0$ depending on $\Mms$ and $E$. The SEDs
and color temperature evolution depend on $\Mms$ (\ie
$\Rs$) and $E$, while their dependencies on $Z$ are small.

The semi-analytic solutions for shock breakout by
\cite{mat99} provide radiation temperature $\Tmm$, outburst energy $\Emm$,
timescale $\tmm$, and luminosity $\Lmm~(=\Emm/\tmm)$ for
polytropic envelope structures; 
the light travel time or limb darkening corrections are not included. In order to compare
our results with the semi-analytical solutions, we extract four following
characteristics of shock breakout from the corrected and uncorrected
models; $\Tp$: color temperature at $t=0$, $\thalf$: days until bolometric
magnitude declines by 1 magnitude after the bolometric peak, $\Erad$:
radiation energy emitted from $t=-\thalf$ to $t=\thalf$, 
and $\Lp$: peak bolometric luminosity. The
properties of our models are summarized in Table~\ref{tab:SB}. Their
dependencies on $\Rs$ and $E$ are shown in Figures~\ref{fig:SB}a-\ref{fig:SB}h 
and compared with the semi-analytic solutions. Since the semi-analytic
solutions slightly depend on $\Mej$ that is different in the numerical models by
a factor of $\sim2$. Possible ranges of the semi-analytic solutions are
shown in Figures~\ref{fig:SB}a-\ref{fig:SB}h.

\begin{figure*}[t]
\epsscale{1.}
\plotone{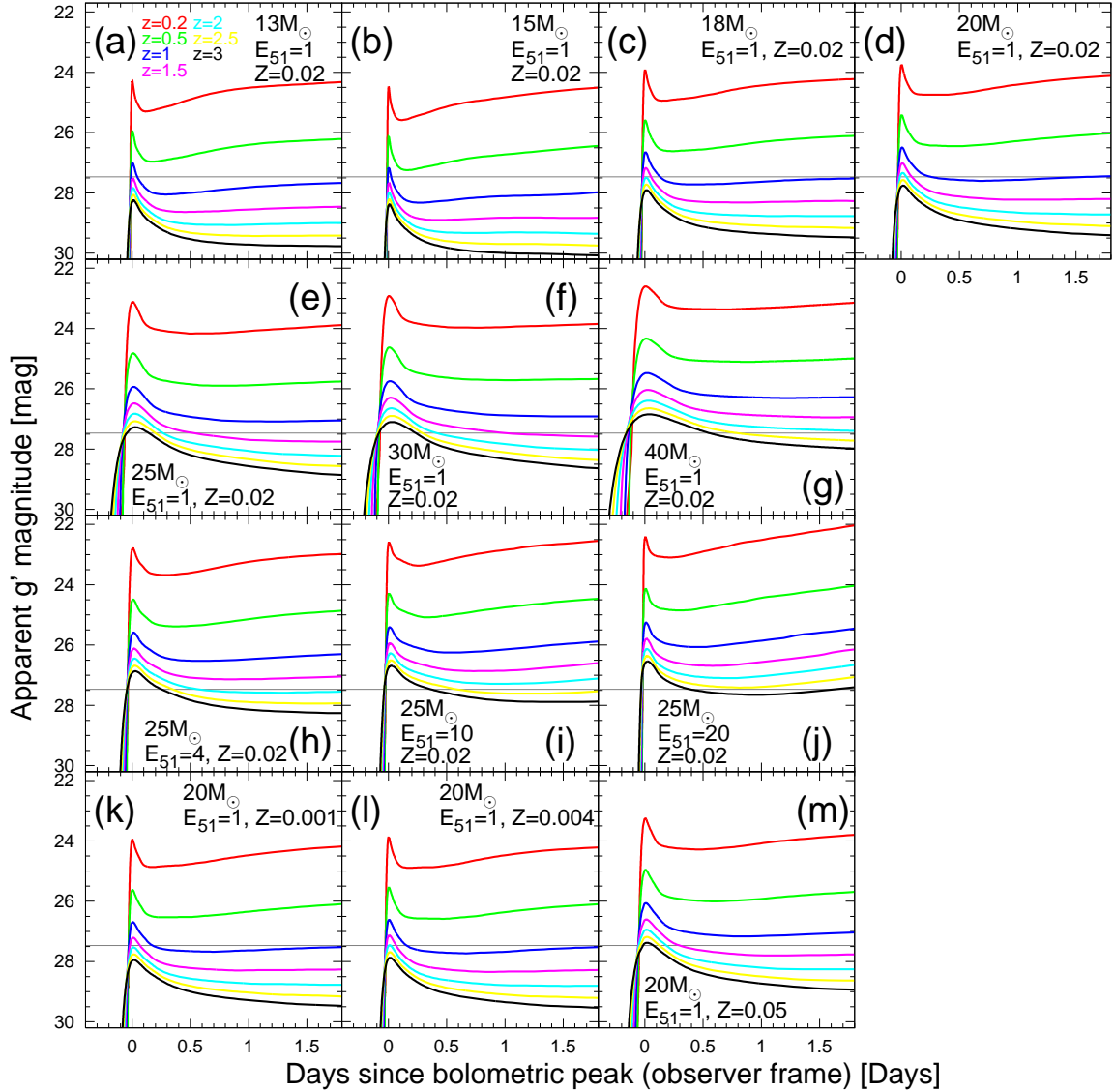} 
\caption{Apparent $g'$-band LCs of the models at 
 $z=0.2$ ({\it red}), $z=0.5$ ({\it green}), $z=1$ ({\it blue}), 
 $z=1.5$ ({\it magenta}), $z=2$ ({\it cyan}), $z=2.5$ ({\it yellow}),
 and $z=3$ ({\it black}). No extinction and no IGM absorption are assumed.
 The panels are the same as in Figure~\ref{fig:AbsLC}.
 The horizontal line shows a 5$\sigma$ detection limit in $g'$ band for
 Subaru/Suprime-Cam 1 hour integration
 ({\it gray}, {http://www.naoj.org/cgi-bin/spcam\_tmp.cgi}, assuming
$0''\hspace{-1ex}.7$ seeing, $1''\hspace{-1ex}.5$ aperture, and 3 days
from New Moon).
}
\label{fig:gLC}
\end{figure*}

Comparing the uncorrected models and the semi-analytic solutions,
they are quantitatively different but the dependencies
are roughly consistent. The models with larger $\Rs$ have lower $\Tp$, longer $\thalf$, and
higher $\Erad$. The models with higher $E$ have higher $\Tp$,
$\Erad$, and $\Lp$ and shorter $\thalf$. The models with different $Z$
are distributed along a sequence of the models with different $\Rs$. This indicates that
the variations with $Z$ can be interpreted by the variation with $\Rs$ and
that the metallicity alters shock breakout mainly through the
variation of stellar structure. Accordingly, the
$Z=0.02$ models could be applied even for shock breakout in stars
with different $Z$ if they have the same $\Rs$.\footnote{It is obviously
better to perform a radiation hydrodynamics calculation 
for an evolutionary presupernova model with proper $Z$.}

The difference between $\Tp$ and $\Tmm$ partly stems from the fact
that the color temperature is different from the radiation
temperature by definition when the opacity depends on frequencies and
the SED is not blackbody.
However, it is notable that the semi-analytic solution is roughly
consistent with the numerical models if $\Tmm$ is reduced by a
factor of 1.5 (Figs.~\ref{fig:SB}a and \ref{fig:SB}b). 
The semi-analytic solutions give slightly higher values also for
the other properties. They are in agreement with the
numerical models if $\tmm$,
$\Emm$, and $\Lmm$ are reduced by a factor of 2, 3, and 1.5, respectively
(Figs.~\ref{fig:SB}c-\ref{fig:SB}h). The reduction factors make it
possible to approximately derive progenitor properties from a direct
comparison between observations and the semi-analytic solutions. Moreover,
the qualitative consistency with the semi-analytic solutions supports the
reliability of our numerical results. 

\begin{figure*}[t]
\plotone{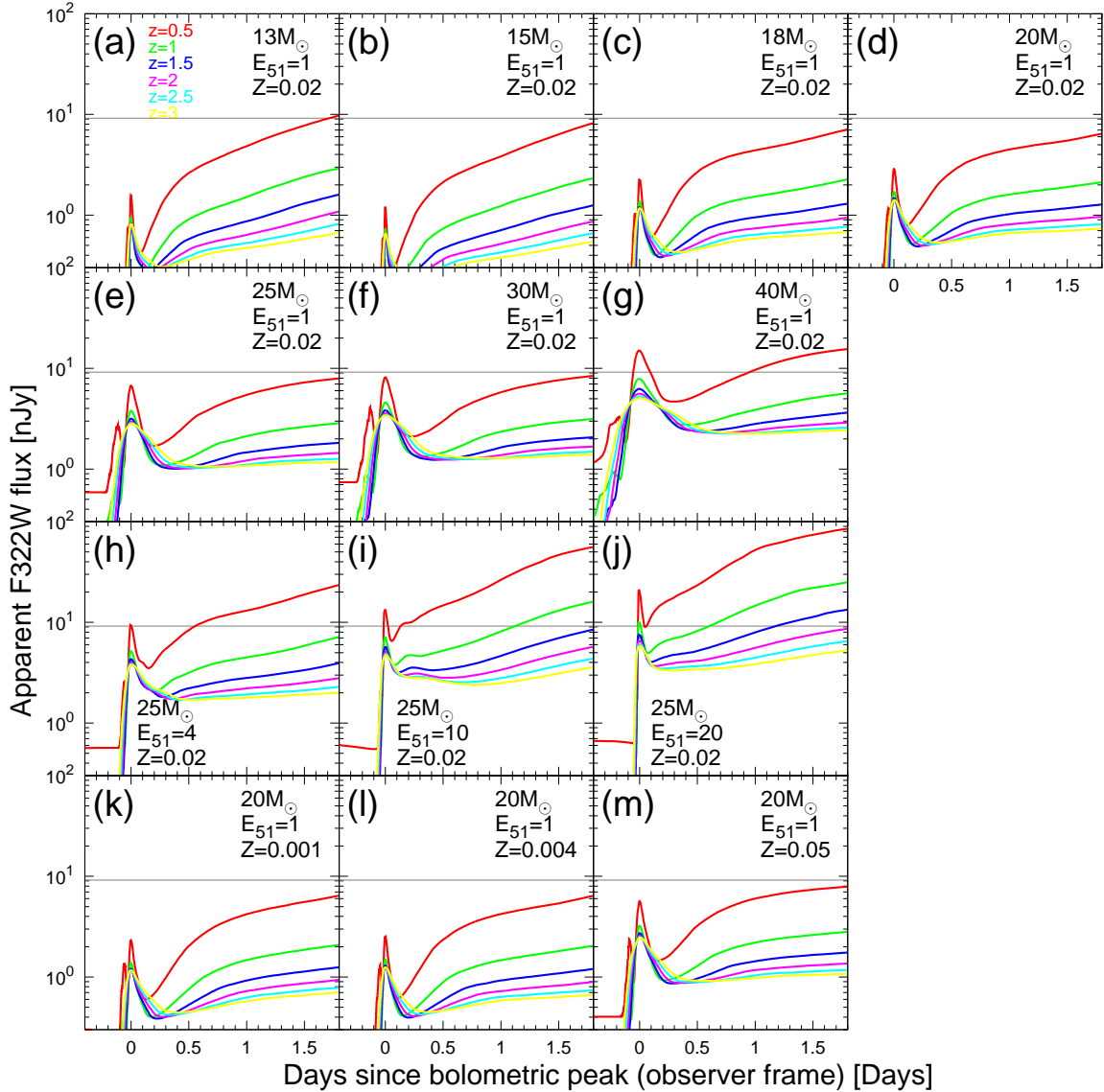} 
\caption{Apparent F322W-band LCs of the models at $z=0.5$ ({\it red}), $z=1$
 ({\it green}), $z=1.5$ ({\it blue}), 
 $z=2$ ({\it magenta}), $z=2.5$ ({\it cyan}), and $z=3$ ({\it yellow}).
 No extinction and no IGM absorption are assumed.
 The panels are the same as in
 Figure~\ref{fig:AbsLC}. The horizontal line shows a 10$\sigma$ detection
 limit in F322W band for JWST/NIRCam $10^4$~sec integration ({\it
 gray}, {http://www.stsci.edu/jwst/instruments/nircam/sensitivity/index\_html}).
}
\label{fig:3.2umLC}
\end{figure*}

The light travel time and limb darkening corrections slightly reduce
$\Tp$ and enhance $\Erad$ but does not change their dependencies. On the
other hand,
the corrections considerably change the dependencies of $\thalf$ on $\Rs$
and $E$, and the dependence of $\Lp$ on $E$. This is because the
corrections smear the LC peak and redistribute radiation energy emitted 
at bright epochs to a time range of $\Rph/c$, where $\Rph$ is a
photospheric radius. Consequently, the corrections lengthen $\thalf$ and diminish
$\Lp$ more efficiently for a model with shorter $\thalf$ and brighter
$\Lp$. Shock breakout in the model with smaller $\Rs$ and
higher $E$ has shorter $\thalf$ and brighter $\Lp$
(Figs.~\ref{fig:SB}c-\ref{fig:SB}d and \ref{fig:SB}g-\ref{fig:SB}h) and
thus more strongly corrected.

\vspace{1cm}

\subsection{Shock breakout at high redshift}
\label{sec:highz}

\begin{figure*}[t]
\plotone{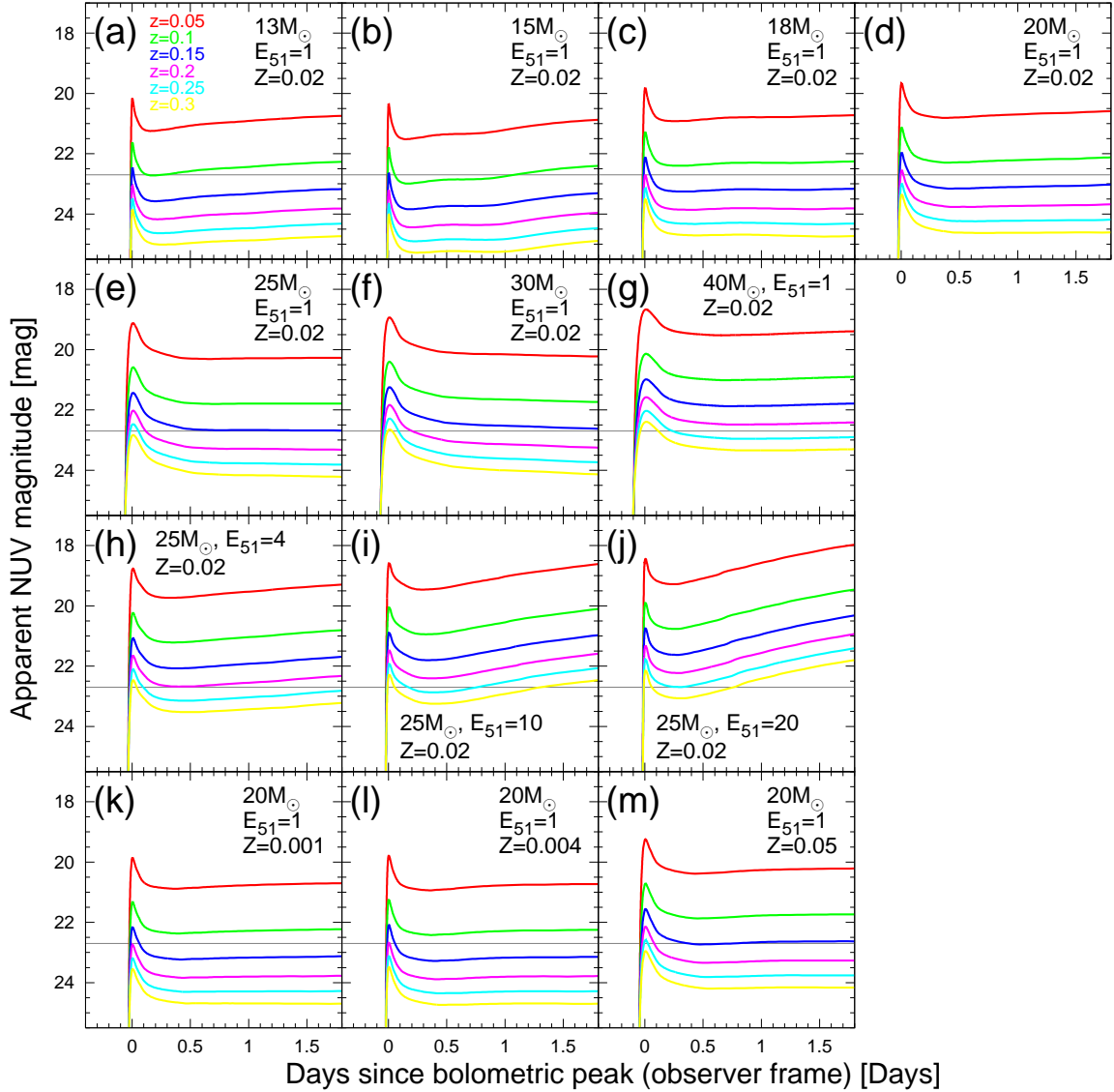} 
\caption{Apparent NUV-band LCs of the models at $z=0.05$ ({\it
 red}), $z=0.1$ ({\it green}), $z=0.15$ ({\it blue}), 
 $z=0.2$ ({\it magenta}), $z=0.25$ ({\it cyan}), and $z=0.3$ ({\it
 yellow}). No extinction and no IGM absorption are assumed.
 The panels are
 the same as in Figure~\ref{fig:AbsLC}. The horizontal line
 shows a 5$\sigma$ detection limit in NUV band for {\it GALEX} satellite 1500s
 integration ({\it gray}, \citealt{mor07}). 
}
\label{fig:NUVLC}
\end{figure*}

\begin{figure*}[t]
\plotone{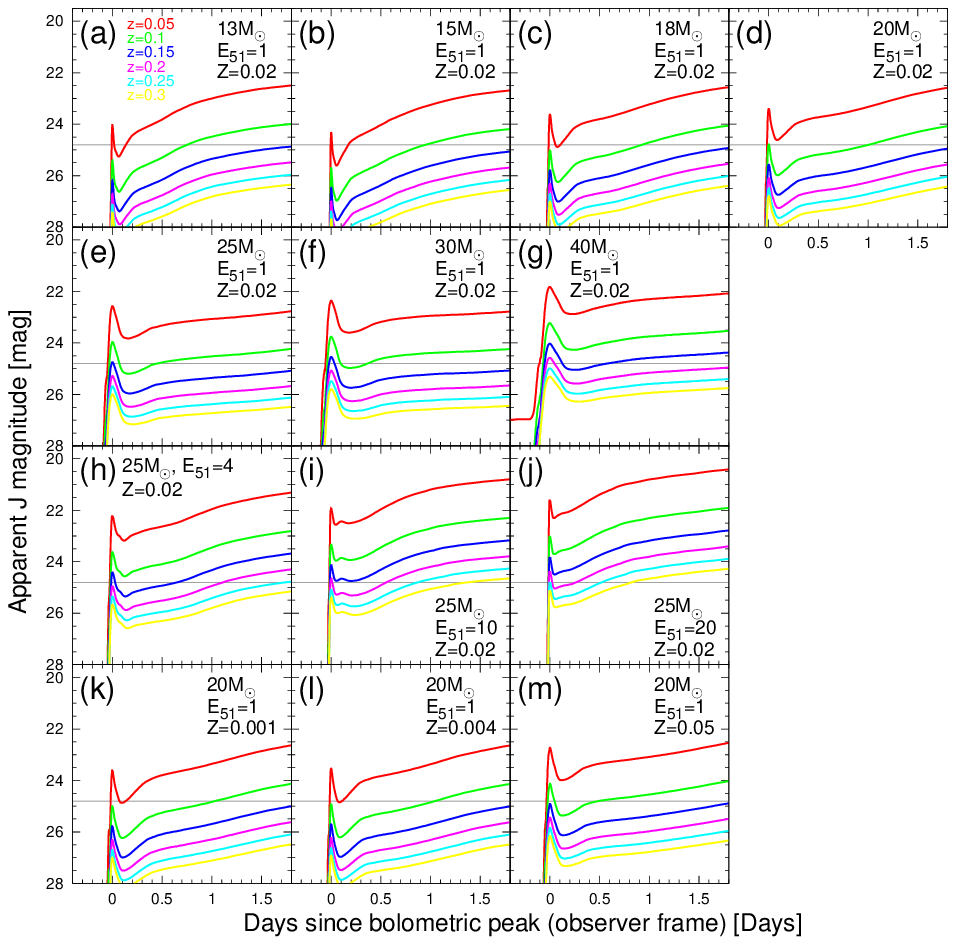} 
\caption{Apparent $J$-band LCs of the models. No extinction and no IGM absorption are assumed. The panels are the same as in
 Figure~\ref{fig:AbsLC} and the color of lines are the same as in
 Figure~\ref{fig:NUVLC}. The horizontal line shows a 5$\sigma$ detection
 limit in $J$ band for VLT/HAWK-I 1 hour integration ({\it gray}, \citealt{kis08}).
}
\label{fig:JLC}
\end{figure*}

The theoretical multigroup LCs allow us to predict apparent LCs of SNe taking place
at arbitrary distance, direction, and host galaxy. Let us consider the
case where we observe an
object at redshift $z$ which emits light with a luminosity per unit
frequency $L_{\nu_0}$ at frequency $\nu_0$ in the rest frame and
we detect the object with a flux per unit frequency $f_{\nu_{\rm obs}}$ 
at frequency $\nu_{\rm obs}$ in the observer frame. Here  
$\nu_{\rm obs}=\nu_0/(1+z)$. The observed flux is obtained as follows:
\begin{equation}
 f_{\nu_{\rm obs}}(\nu_{\rm obs})= {1\over{4\pi D_{\rm L}^2}} (1+z) L_{\nu_0}\left({\nu_0}\right)
\end{equation}
where $D_{\rm L}$ is the luminosity distance. Here, we adopt for
cosmological parameters a five-year
result of {\it Wilkinson Microwave Anisotropy Probe} \citep{kom09}: 
$H_0=70.5~{\rm km~s^{-1}~Mpc^{-1}}$, $k=0$, $\Omega_\lambda=0.726$,
and $\Omega_{\rm M}=0.274$. Apparent multicolor LCs are derived by
convolving the diluted and redshifted multigroup LCs with the
bandpasses of the satellites or telescopes.

The LCs in $g'$ band for $z=0.2$, $0.5$, $1$, $1.5$, $2$, $2.5$,
and $3$, LCs in F322W band for $z=0.5$, $1$, $1.5$, $2$, $2.5$,
and $3$, and LCs in NUV and $J$ bands for $z=0.05$, $0.1$, $0.15$, $0.2$,
$0.25$, and $0.3$ are shown in Figures~\ref{fig:gLC}a-\ref{fig:gLC}m,
\ref{fig:3.2umLC}a-\ref{fig:3.2umLC}m, \ref{fig:NUVLC}a-\ref{fig:NUVLC}m, 
and \ref{fig:JLC}a-\ref{fig:JLC}m, respectively. 
These figures also show the limiting magnitudes of telescopes/instruments
with wide-field imaging capability: {\it GALEX} satellite
in NUV band for $5\sigma$ detection with $1500$~sec integration ($m_{\rm NUV,lim}=22.7$~mag,
\citealt{mor05,mor07}, Figs.~\ref{fig:NUVLC}a-\ref{fig:NUVLC}m), Subaru/Suprime-Cam in $g'$ band for $5\sigma$
detection with 1 hour integration
($m_{g',{\rm lim}}=27.5$~mag,\footnote{The limiting magnitude is calculated
with Subaru Imaging Exposure Time Calculator
(\url{http://www.naoj.org/cgi-bin/spcam\_tmp.cgi}) assuming
$0''\hspace{-1ex}.7$ seeing, $1''\hspace{-1ex}.5$ aperture, and 3 days
from New Moon.} \citealt{miy02}, Figs.~\ref{fig:gLC}a-\ref{fig:gLC}m), VLT/HAWK-I in $J$ band for $5\sigma$ detection with 1 hour integration
($m_{J,{\rm lim}}=24.8$~mag, \citealt{kis08}, Figs.~\ref{fig:JLC}a-\ref{fig:JLC}m), and
{\it JWST}/Near Infrared Camera (NIRCam) in
F322W band for $10\sigma$ detection with $10^4$~sec integration
($9.18$~nJy,\footnote{\url{http://www.stsci.edu/jwst/instruments/nircam/sensitivity/index\_html}.}
Figs.~\ref{fig:3.2umLC}a-\ref{fig:3.2umLC}m).
No extinction and no intergalactic medium (IGM)
absorption are adopted here.

\begin{figure*}
\plotone{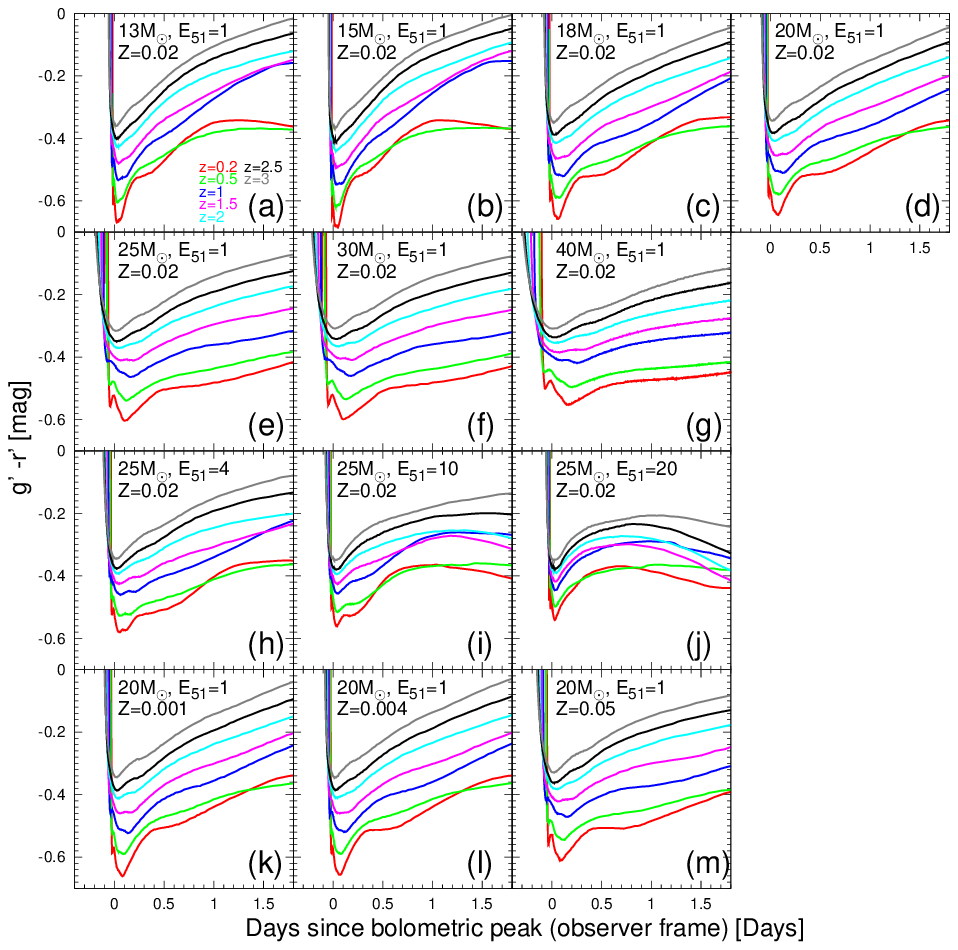} 
\caption{Color ($g'-r'$) evolution of models in the observer frame at 
 $z=0.2$ ({\it red}), $z=0.5$ ({\it green}), $z=1$ ({\it blue}), 
 $z=1.5$ ({\it magenta}), $z=2$ ({\it cyan}), $z=2.5$ ({\it black}),
 and $z=3$ ({\it gray}). 
 The panels are the same as in Figure~\ref{fig:AbsLC}.
}
\label{fig:gr}
\end{figure*}

These figures demonstrate that the shock breakout with higher $E$ or larger $\Rs$,
\ie larger $\Mms$ or higher $Z$, can be detected at higher redshift. 
The shock breakout can be detected in $g'$ band even at $z\sim1$
(13 and 15~$\Msun$ models), $z\sim2$ (18 and 20~$\Msun$ models), and
$z\sim3$ (25, 30, and 40~$\Msun$). On the
other hand, the observations in NUV and near infrared (NIR) bands can
detect shock breakout only at $z\lsim0.5$, although SNe at the later
epoch, \ie plateau stage, are detectable even at $z\gsim4$ by {\it JWST}
(N.~Tominaga, \etal\ in prep.). This is because
the limiting magnitude in UV bands is much shallower than that in optical
bands and the SED of shock breakout is too blue for the NIR
observations. The shock breakout
has a blue color in optical (Figs.~\ref{fig:gr}a-\ref{fig:gr}m), hence
an observation in bluer optical bands is more suitable to detect
shock breakout as long as the
IGM absorption is irrelevant in the adopted bandpass.

\begin{figure*}
\epsscale{1.}
\plotone{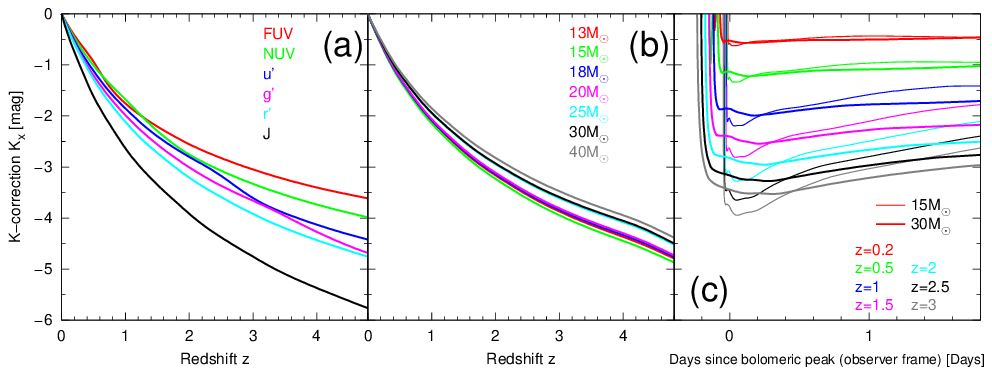} 
\caption{$K$-corrections $K_x$ between rest-frame $x$ band and observer-frame
 $x$ band. (a)  $K_{\rm FUV}$ ({\it red}), $K_{\rm NUV}$
 ({\it green}), $K_{u'}$ ({\it blue}), $K_{g'}$ ({\it magenta}), $K_{r'}$
 ({\it cyan}), and $K_J$ ({\it black}) of the $20\Msun$,
 $Z=0.02$, and $E_{51}=1$ model as a function of redshift.
(b) $K_{g'}$ of the $Z=0.02$ and
 $E_{51}=1$ models with
 $\Mms=13\Msun$ ({\it red}), $15\Msun$ ({\it green}), $18\Msun$ ({\it
 blue}), $20\Msun$ ({\it magenta}), $25\Msun$ ({\it cyan}), $30\Msun$
 ({\it black}), and $40\Msun$ ({\it gray}) as a function of redshift.
(c) $K_{g'}$ of the $Z=0.02$ and
 $E_{51}=1$ models with $\Mms=15\Msun$ ({\it thin lines}) and $30\Msun$
 ({\it thick lines}) at $z=0.2$ ({\it red}), $0.5$
 ({\it green}), $1$ ({\it blue}), $1.5$ ({\it magenta}), $2$ ({\it
 cyan}), $2.5$ ({\it black}), and $3$ ({\it gray}) as a function of time
 in the observer frame.
}
\label{fig:Kcor}
\end{figure*}

Shock breakout has large negative $K$-corrections between
rest-frame $x$ band and observer-frame $x$ band $K_x$, where $x$ band is 
an arbitrary bandpass at $\lambda>100$~\AA.
Figure~\ref{fig:Kcor}a shows $K_{\rm FUV}$, $K_{\rm NUV}$, $K_{u'}$,
$K_{g'}$, $K_{r'}$, and $K_J$ of the $20\Msun$, $Z=0.02$, and
$E_{51}=1$ model at $t=0$. Since the SED at $t=0$ is extremely blue, the
negative $K$-corrections are larger for higher redshift even in FUV band and the
$K$-corrections in redder bands are smaller than those in bluer bands. 
Figure~\ref{fig:Kcor}b shows $K_{g'}$ of the $Z=0.02$ models
with different $\Mms$ at $t=0$. The more massive
models have slightly larger $K$-corrections by $\lsim0.5$~mag
at $z\lsim4.5$. Figure~\ref{fig:Kcor}c shows the evolution of $K_{g'}$
of the $15\Msun$ and $30\Msun$ models. The
$K$-correction evolves more rapidly for higher redshift or smaller
$\Mms$. As a result, $K_{g'}$ of the $15\Msun$ model is
smaller at $t=0$ but larger at $\tobs=1.5$~days than that of the $30\Msun$
model, where $\tobs$ is a time from $t=0$ in the observer frame. 

\begin{figure*}
\epsscale{1.}
\plotone{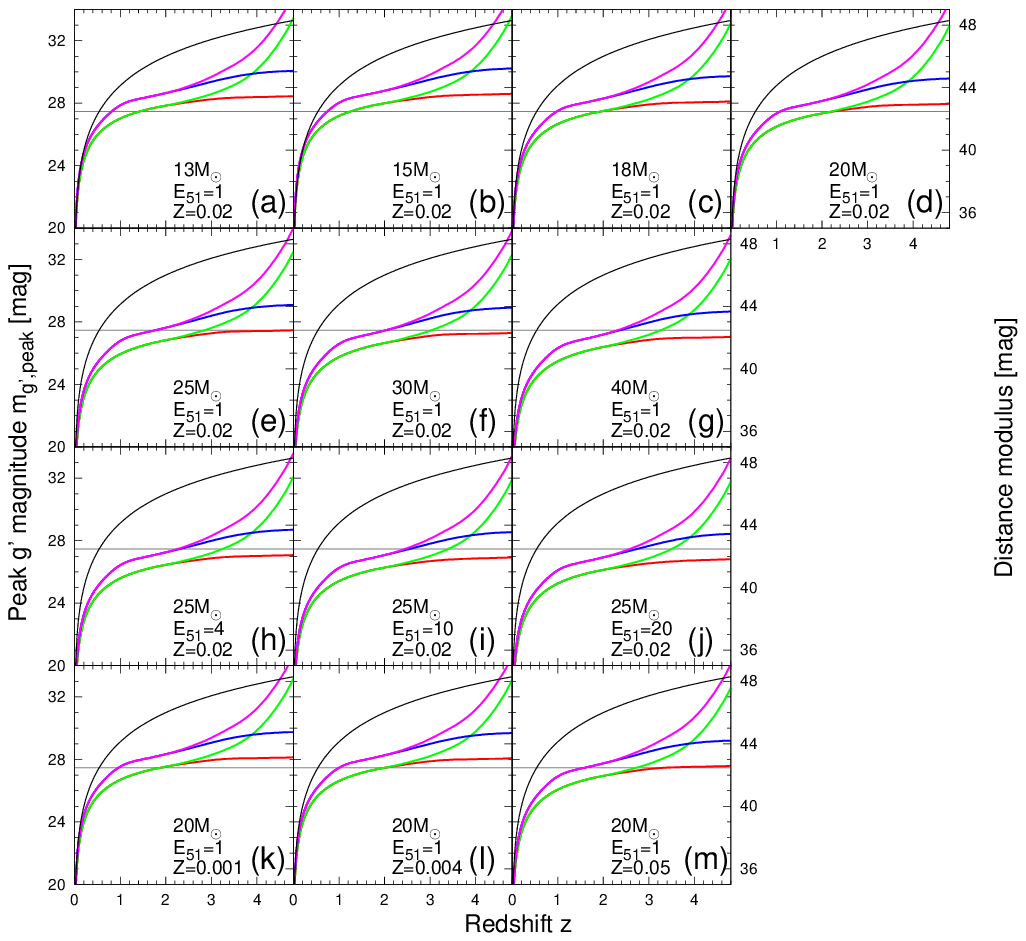} 
\caption{Peak $g'$ magnitude $m_{g',{\rm peak}}$ and distance modulus
 ({\it black}) as a function of redshift. The peak magnitudes are
 derived by assuming no extinction and IGM absorption ({\it red}), host
 galaxy extinction ($\Ebvh=0.1$~mag and our Galaxy extinction law,
 \citealt{pei92}) ({\it blue}), IGM absorption (\citealt{mad95}) ({\it
 green}), and host galaxy extinction ($\Ebvh=0.1$~mag and our Galaxy
 extinction law) and IGM absorption ({\it magenta}). The
 panels are the same as in Figure~\ref{fig:AbsLC} and the horizontal
 line ({\it gray}) is the same as in Figure~\ref{fig:gLC}.
}
\label{fig:peakz}
\end{figure*}

Figures~\ref{fig:peakz}a-\ref{fig:peakz}m show distance modulus and
apparent peak $g'$-band magnitudes $m_{g',{\rm peak}}$ of models as a function
of redshift for different assumptions on the host galaxy extinction and
IGM absorption \citep{mad95}. Here, we assume that the host galaxy has a
color excess as $\Ebvh=0$ or $0.1$~mag and our Galaxy extinction law
\citep{pei92}. These figures display that the dimming of apparent
magnitude is considerably weak compared to the distance modulus
because of the large negative $K$-correction. For illustration purpose,
a 1-hour limiting magnitude for Subaru/Suprime-Cam is also shown
($\mglim=27.5$~mag). The maximum redshift for detecting shock breakout
with 8m-class telescopes is mainly determined by the host galaxy extinction,
while the IGM absorption becomes relevant at $z\gsim2$ in $g'$ band. 
When a limiting magnitude $m_{g',{\rm lim}}\sim30$~mag is achieved by
next-generation $30$m-class telescopes, the maximum redshifts 
are $z\sim3.5$ for $\Ebvh=0.1$~mag and $z\sim4.2$
for $\Ebvh=0$ which are limited by the IGM absorption. 
At that era, shock breakout can be used for studies on not only distant
SNe~II-P, host galaxy extinction, or evolution of universe but also an IGM.

\subsection{Future prospects for shock breakout surveys}
\label{sec:future}

The detection of shock breakout has recently been consummated but the number is still small.
This is because it is difficult for past/ongoing SN/transient surveys with several
days intervals to detect short-term soft X-ray/UV flashes. Hence, we
propose a deep optical survey with short
intervals, \eg $\sim$~hour, for shock breakout in an SN~II-P in the distant
universe.

This is motivated by the following two brand-new prospects. (1) Due to
the large negative $K$-correction, distant shock breakout up to 
$z\sim3$ is bright enough to be detected with current optical
facilities (\S~\ref{sec:highz}). Such a survey is promising
because of three reasons: the duration
of a distant event is elongated, the star formation rate (SFR) is high in
the distant universe (\eg \citealt{hop06}), and available optical facilities
are numerous compared to X-ray/UV satellites.
(2) The multiepoch imaging observation in a night is essential to draw
the LCs of shock breakout in both rising and declining phases because the time scale
of shock breakout in an SN~II-P at $z\lsim3$ are less than $\sim1$ day
in the observer frame (\S~\ref{sec:highz}).

In the following, we estimate the expected number and highest
redshift of detection, discuss influences of uncertainties on host
galaxy extinction and SFH, and propose realistic and promising survey
strategies, ways to identify shock breakout, and ways to constrain the
SN properties from observable quantities.
In this section, we focus on the
models with $E_{51}=1$ and $Z=0.02$.

\subsubsection{Expected number}
\label{sec:number}

Host galaxy extinction heavily reduces brightness of shock breakout
(\S~\ref{sec:highz}), and thus it should be taken into account for a
realistic number estimate. However, the
host galaxy extinction of distant SNe~II-P is unknown but instead it
will be clarified by future shock breakout studies.
Hence we expediently assume that the distribution of host galaxy extinction
of distant SNe~II-P is equivalent to that of nearby SNe II-P.
The host galaxy extinction of a nearby SN~II-P is estimated from \ion{Na}{1}-D
lines of the host galaxy, a spectroscopic observation of SNe, or a
color of SN plateau (\eg \citealt{kri09,oli10}). We employ the
distribution of host galaxy extinction presented in \cite{oli10}. Although we take
conservatively the highest extinction for each SN among their estimates, we
caution that it could be biased towards bright, \ie less-reddened,
SNe~II-P.

We estimate the incidence rate of shock breakout that
can be brighter than a limiting magnitude $\mlim$ in bandpass $x$, named
``observable SN rate''. The procedure to estimate observable SN rate is
as follows: we define a peak magnitude in bandpass $x$ 
[$m_{{\rm peak},x}(\Mms,A_{\rm V},z)$] for an SN with $\Mms$ which explodes in
a galaxy with extinction $A_{\rm V}$ at a redshift $z$, and then
a detection probability $f_x[m_{{\rm peak},x}(\Mms,A_{\rm V},z),\mlim]$
is set to $1$ when $m_{{\rm peak},x}(\Mms,A_{\rm V},z)\leq\mlim$ or 
$0$ when $m_{{\rm peak},x}(\Mms,A_{\rm V},z)>\mlim$. Accordingly, the
observable SN rate per unit solid angle per unit time in the observer
frame $n_x(\mlim)$ is obtained by integrating the detection probability
with an IMF $\phi(\Mms)$, a cosmic SFH $\eta(z)$, and the
distribution of host galaxy extinction 
$\chi(A_{\rm V})$, where $\int\chi(A_{\rm V})dA_{\rm V}=1$, as 
\begin{eqnarray}
 n_x(\mlim) = \iiint f_x[m_{{\rm peak},x}(\Mms,A_{\rm V},z),\mlim]
  \nonumber \\
  \phi(\Mms) {dV(z)\over{d\Omega dz}}
 {\eta(z)\over{1+z}} \chi(A_{\rm V}) d\Mms dz dA_{\rm V},
\label{eq:nx}
\end{eqnarray}
where $V(z)$ is a comoving volume up to $z$ and $\Omega$ is a solid
angle. Here, we assume a modified Salpeter A IMF (\citealt{sal55,bal03})
and a cosmic SFH in \cite{hop06}, and we 
adopt approximate formulae for the IGM absorption \citep{mad95} and the
host galaxy extinction (our Galaxy extinction law,
\citealt{pei92}).\footnote{Since these formulae cannot be
adopted at $\lambda<800$~\AA\ in the rest frame, we assume that the absorption and
extinction at $\lambda<800$~\AA\ are the same as those at
$\lambda=800$~\AA.} 

\begin{figure}[t]
\epsscale{1.}
\plotone{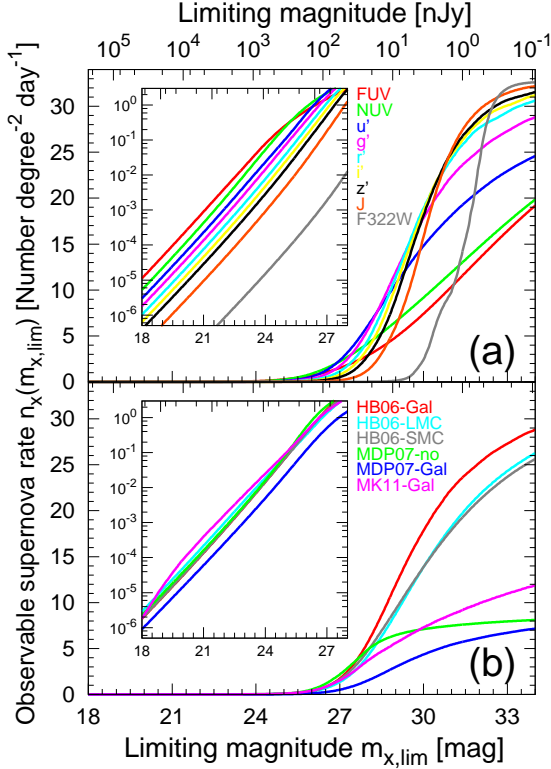} 
\caption{Observable SN rates per square degree per day $n_x(\mlim)$ for
 (a) different bandpass: FUV ({\it red}),
 NUV ({\it green}), $u'$ ({\it blue}), $g'$ ({\it magenta}),  $r'$ ({\it cyan}), $i'$ ({\it
 yellow}), $z'$ ({\it black}), $J$ ({\it orange}), and F322W ({\it
 gray}) and for (b) different host galaxy extinction and SFH in $g'$ band:
 estimates HB06-Gal ({\it red}), HB06-LMC ({\it cyan}), HB06-SMC ({\it
 gray}), MDP07-no ({\it green}), MDP07-Gal ({\it blue}), and MK11-Gal
 ({\it magenta}). The insets enlarge the regions with $\mlim\leq28$~mag.
}
\label{fig:magnum}
\end{figure}

The observable
SN rate per square degree per day are shown as a function of $\mlim$ in
Figures~\ref{fig:magnum}a and \ref{fig:magnum}b. 
For deeper limiting magnitude, the observable SN rates are higher due to the larger
observable volume and the higher SFR at high redshift. For example,
the $g'$-band observations with $\mglim=20$, $22$, $24$, $26$, $27$, $28$, and
$30$~mag detect $3.1\times10^{-5}$, $6.2\times10^{-4}$,
$1.4\times10^{-2}$, $3.7\times10^{-1}$, $1.7$,
$5.6$, and $1.8\times10$~${\rm SNe~degree^{-2}~day^{-1}}$,
respectively. $n_{g'}(\mglim)$ logarithmically increases with $\mglim$
at $\mglim\lsim27.5$~mag, but the increase slows down at
$\mglim\gsim27.5$~mag because the maximum redshift of observable SNe
reaches $z\sim2.5$ (Figs.~\ref{fig:peakz}a-\ref{fig:peakz}m) where the
IGM absorption becomes nonnegligible and the cosmic SFH hits a peak.

\begin{figure}[t]
\epsscale{1.}
\plotone{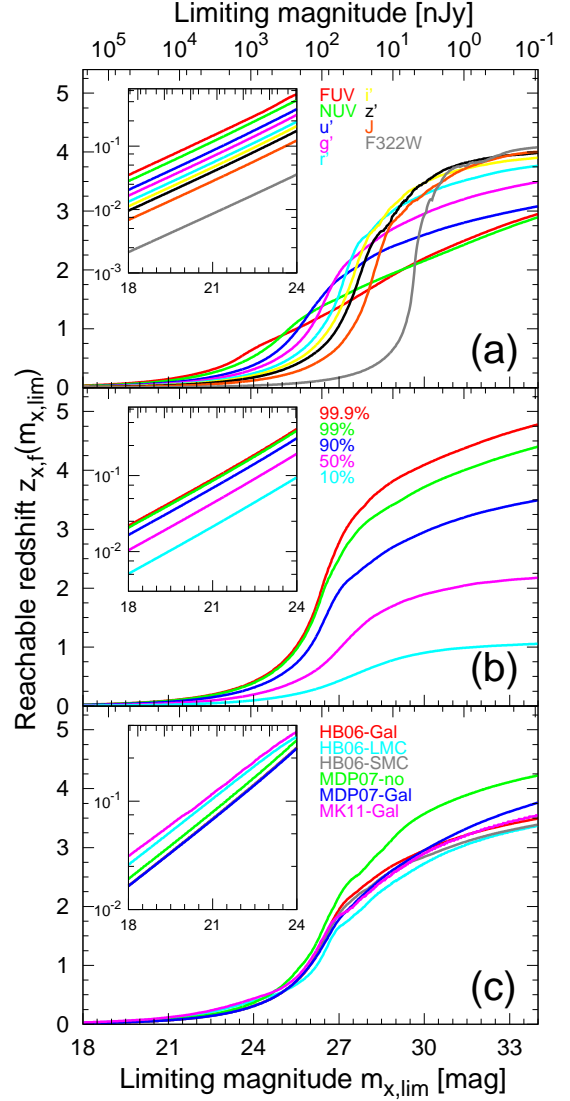} 
\caption{Reachable redshifts $\zlim(\mlim)$ for (a) different bandpass with $f=0.9$: the
 colors are the same as in Fig.~\ref{fig:magnum}a, for (b) different
 $f$ in $g'$ band: $f=0.999$ ({\it red}), $0.99$ ({\it green}),
 $0.9$ ({\it blue}), $0.5$ ({\it magenta}), and $0.1$ ({\it cyan}),
 and for (c) different host galaxy extinction and SFH in $g'$ band; the colors are the same as in
 Fig.~\ref{fig:magnum}b. The insets enlarge the regions with $\mlim\leq24$~mag.
}
\label{fig:highz}
\end{figure}

\begin{figure}[t]
\plotone{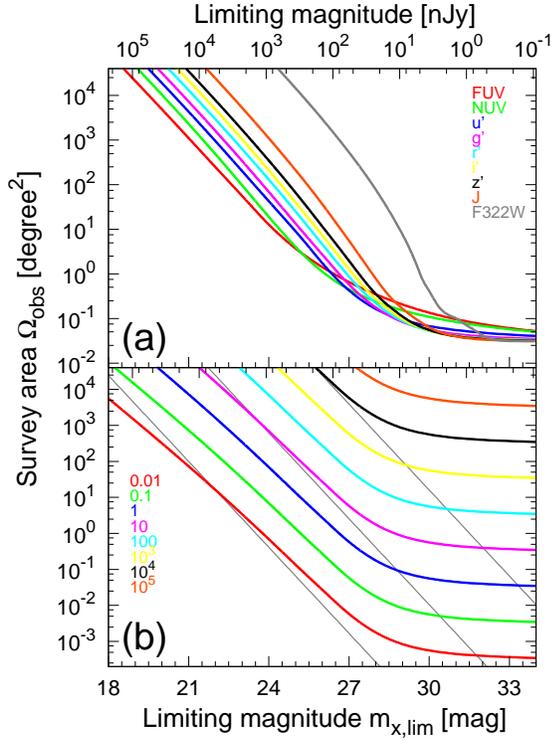} 
\caption{Observable SN rate per day $\Omega_{\rm obs} n_x(\mlim)$ as a function of
 limiting magnitude $\mlim$ and survey area $\Omega_{\rm obs}$. 
 (a) $\Omega_{\rm obs} n_x(\mlim)=1$ in different bandpass: the
 colors are the same as in Fig.~\ref{fig:magnum}a and (b) different
 $\Omega_{\rm obs} n_x(\mlim)$ in $g'$ band: 
 $\Omega_{\rm obs} n_x(\mlim)=0.01$ ({\it red}), $0.1$ ({\it green}), $1$
 ({\it blue}), $10$ ({\it magenta}), $100$ ({\it cyan}), $10^3$ ({\it
 yellow}), $10^4$ ({\it black}), and $10^5$ ({\it orange}). The
 lines being accessible by equal survey powers are shown ({\it
 gray}).
}
\label{fig:depwid}
\end{figure}

Redshifts below which a given fraction $f$ of observable SNe locates
[$\zlim(\mlim)$, named ``reachable redshift''] are shown in
Figures~\ref{fig:highz}a-\ref{fig:highz}c. 
The fraction of high-$z$ events increases with $\mlim$. For example,
the $g'$-band observations with $\mglim=20$, $22$, $24$, $26$, $27$, $28$, and
$30$~mag result in [$z_{g',0.1}(\mglim)$, $z_{g',0.5}(\mglim)$, 
$z_{g',0.9}(\mglim)$] = ($0.013$, $0.027$, $0.042$), ($0.035$, $0.071$,
$0.11$), ($0.095$, $0.19$, $0.31$), ($0.26$, $0.55$, $1.0$), ($0.42$,
$0.96$, $1.9$), ($0.62$, $1.4$, $2.4$), and ($0.90$, $1.9$, $3.0$),
respectively. 
If $\mglim\geq27$~mag is realized, more than half of the observable
SNe take place at $z\geq0.9$. This capability to access high-$z$
universe is an intriguing and unique feature of shock breakout.

If the same
limiting magnitude is available, $n_x(\mlim)$ and $\zlim(\mlim)$ in
bluer bands are higher than those in redder bands at shallow $\mlim$, \eg
$\mlim\lsim24-25$~mag for UV, due to the blue SED of shock
breakout. However, the increases in $n_x(\mlim)$ and  $\zlim(\mlim)$
for deep $\mlim$ are suppressed by the
IGM absorption. Therefore, $n_x(\mlim)$ and $\zlim(\mlim)$ in
redder bands overcome those in bluer bands for deep $\mlim$. Thus, the
most effective bandpass for shock breakout detection depends on
feasible $\mlim$. For example, the $g'$- and $r'$-band observations are
currently the most effective in number and in reachable redshift,
respectively, because only 8m-class optical telescopes or Hubble Space
Telescope ($\mlim\lsim28$~mag) are
available at the moment. When 30m-class optical/infrared telescopes
($\mlim\sim30$~mag) become operative, the most effective bands will be
$r'$ band in number and $i'$ band in reachable redshift.

The reachable redshift dramatically increases if $\mlim$ is
deeper than $26-30$~mag (Figs.~\ref{fig:highz}a-\ref{fig:highz}c) because the large
negative $K$-correction makes the apparent peak magnitudes of shock
breakout almost constant ($\sim26-30$~mag) for a wide redshift range
(Figs.~\ref{fig:peakz}a-\ref{fig:peakz}m). Therefore, in order to detect
distant shock breakout at $z\gsim1$, it is essential to attain
$\mlim\gsim26-30$~mag. Since the dramatical increases of reachable
redshift in optical bands coincide with the
limiting magnitudes of current optical facilities, the improvement
in near future will enhance the reachable redshift considerably. 
On the other hand,
the increase in $z_{x,0.9}(\mlim)$ reaches the ceiling at $z\sim4$ for
$\mlim\gsim30$~mag even in NIR bands. This is because the cosmic SFH in
\cite{hop06} has low SFR at such a high redshift. It is important to
note that the detection of shock breakout at $z>4$ is feasible
if an SFR is high enough, unless the bandpass are below the rest-frame
Ly$\alpha$ wavelength.

According to $n_x(\mlim)$, observable SN rates per unit time 
$\Omega_{\rm obs} n_x(\mlim)$ with a given $\mlim$ and a survey area
$\Omega_{\rm obs}$
are shown in Figures~\ref{fig:depwid}a and \ref{fig:depwid}b. An observation in a
bluer band is more efficient with the same $\mlim$ if a wide survey area is available
($>1$~degree$^2$), while an observation in a redder band is
slightly better if only a narrow survey area observation
($<0.1$~degree$^2$) with deep $\mlim$ ($\gsim30$~mag) is available. The
flat dependence of $\Omega_{\rm obs} n_x(\mlim)$ on $\mlim$ at deep $\mlim$, \eg
$\mglim\gsim28$~mag in $g'$ band, stems from the suppression of
$n_x(\mlim)$ by the IGM absorption and the low SFR at high redshift. 

Figure~\ref{fig:depwid}b shows lines giving equal $\Omega_{\rm obs} n_{g'}(\mglim)$
with the $g'$-band observation. The {\it gray} lines represent equal
survey powers without taking into account observing overhead such as
readout time. The
number of observable SNe is larger for wider and shallower observations
with a given survey power. However,
such a wide and shallow observation misses high-$z$ events
(Figs.~\ref{fig:highz}a-\ref{fig:highz}c) and 
$\Omega_{\rm obs} n_{g'}(\mglim)$ has an upper limit somewhere if the
overhead is taken into account. Therefore, the practical survey parameters
should be customized to purposes of observations and adopted
telescopes/instruments, considering the number, reachable redshift, and overhead. 

\subsubsection{Dependencies on host galaxy extinction and star
   formation history}
\label{sec:uncertain}

We set the above estimate with the SFH in \cite{hop06} and our Galaxy
extinction law (HB06-Gal) as a control estimate and investigate the
dependence of $n_x(\mlim)$ on uncertainties of
host galaxy extinction and SFH (Fig.~\ref{fig:magnum}b). 

We attempt Large Magellanic Cloud and Small
Magellanic Cloud extinction laws (estimates HB06-LMC and HB06-SMC,
\citealt{pei92}) for host galaxies. The LMC and SMC extinction
laws have larger absorption at $\lambda<2000$~\AA\ and smaller at
$\lambda\sim2200$~\AA\ in the rest frame than our Galaxy extinction
law. As a result, $n_{g'}(\mglim)$ of estimates HB06-LMC and HB06-SMC are slightly smaller
at $\mglim\gsim27$~mag than that of estimate HB06-Gal but the estimates
HB06-Gal, HB06-LMC, and HB06-SMC are similar at $\mglim\lsim26$~mag. 

\begin{deluxetable*}{cccc|c|cc|c}[t]
 \tabletypesize{\scriptsize}
 \tablecaption{Simulated survey strategies. \label{tab:num}}
 \tablewidth{0pt}
 \tablehead{
   \colhead{Strategy}
 & \colhead{$N_{\rm night}$}
 & \colhead{$n_{\rm obs}$}
 & \colhead{$t_{\rm exp}$}
 & \colhead{$\mlim$\tablenotemark{a}}
 & \colhead{Number\tablenotemark{b}}
 & \colhead{Reachable redshift\tablenotemark{c}}
 & \colhead{Number with overhead\tablenotemark{d}}\\
   \colhead{}
 & \colhead{[nights]}
 & \colhead{[night$^{-1}$]}
 & \colhead{[min]}
 & \colhead{[mag]}
 & \colhead{[degree$^{-2}$]}
 & \colhead{}
 & \colhead{[degree$^{-2}$]}
 }
\startdata
    A&    1&    3&  120& 28.4&    1.56&    2.40&    1.42\\
    B&    2&    3&   60& 28.0&    2.62&    2.21&    2.38\\
    C&    4&    3&   30& 27.6&    3.57&    1.99&    3.25\\
    D&   12&    3&   10& 27.1&    4.66&    1.46&    4.23\\
    E&   24&    3&    5& 26.7&    5.22&    1.07&    4.75\\
    F&   40&    3&    3& 26.4&    5.49&    0.84&    4.70\\
    G&   60&    3&    2& 26.2&    5.81&    0.71&    4.65\\
    H&    2&    4&   45& 27.9&    2.71&    2.25&    2.46\\
    I&    2&    5&   36& 27.8&    2.73&    2.25&    2.49\\
    J&    2&    6&   30& 27.6&    2.71&    2.25&    2.46\\
    K&    2&   12&   15& 27.3&    2.31&    2.12&    2.10\\
    L&    3&    1&  120& 28.4&  0.0651&    2.44&  0.0592\\
    M&    2&    3&  120& 28.4&    3.91&    2.35&    3.55\\
    N&    3&    3&  120& 28.4&    6.26&    2.34&    5.70
\enddata

\tablenotetext{a}{The limiting magnitude is calculated
with Subaru Imaging Exposure Time Calculator
(\url{http://www.naoj.org/cgi-bin/spcam\_tmp.cgi}) assuming
$0''\hspace{-1ex}.7$ seeing, $1''\hspace{-1ex}.5$ aperture, and 3 days
from New Moon.}

\tablenotetext{b}{The number of observable SNe in total, assuming no overhead.}

\tablenotetext{d}{The redshift below which $90\%$ of observable SNe occur.}

\tablenotetext{d}{The number of observable SNe reduced in proportion to
 the overhead, $10\%$ for
 $t_{\rm exp}\geq5$~min and $30$~sec for $t_{\rm exp}<5$~min.}

\end{deluxetable*}

The SFH in \cite{hop06} is derived by scaling UV SFH so as to be
consistent with infrared SFH and thus presumably includes both
visible and dust-obscured star formation. Although we correct the dust
extinction in host galaxies, it could underestimate the host galaxy
extinction. Additionally, some studies suggest that \cite{hop06}
overcorrects the dust extinction and overestimates the SFR by a factor
of $\sim2$ at $z\sim2-3$ (\eg
\citealt{nag04,nag05,bau05,lac10}).
Hence, we test with UV dust-unobscured SFH \citep{man07}.
Although their estimate is limited at $z<2$, the SFR at $z>2$ is assumed
to be the same as the SFR at $z=2$. Since the
dust attenuation of shock breakout brightness in a host galaxy would
be somehow involved in the UV
dust-unobscured SFH, we present an estimate with no additional host galaxy extinction
(estimate MDP07-no) and also test an estimate with extinction correction
with our Galaxy extinction law as a lower limit (estimate
MDP07-Gal). 

Furthermore, we test an SFH and host galaxy extinction
derived from a cold dark matter-based semi-analytic model 
(M.~A.R.~Kobayashi, \etal\ 2011 in prep., see also
\citealt{nag04,kobm07,kobm10})\footnote{The model well explains
observations of nearby and distant galaxy evolutions and dust-unobscured
luminosity density. Since the intrinsic SFR and dust
extinction of each galaxy are provided, the SFH estimate is free from
uncertainties of extinction correction and conversion from the galaxy
luminosity to the SFR.}
and our Galaxy extinction law (estimate MK11-Gal). The estimate MK11-Gal provides
a self-consistent observable SN rate with respect to the host galaxy
extinction and SFR, although the metallicity evolution is ignored in
this paper. The estimates HB06-Gal, MDP07-no, and MK11-Gal are
consistent but they are higher by a factor of $2.5$ than the estimate
MDP07-Gal at $\mglim\leq28$~mag. On the other hand, the differences between the
estimate HB06-Gal and the estimates MK11-Gal and MDP07-no are as large as a factor
of $2-2.5$ at $\mglim\sim30$~mag. Therefore the number count of shock
breakout with 30m-class telescope can constrain how high the SFR is at
high redshift, being independent of galaxy studies.

We also investigate the dependencies of $z_{g',f}(\mglim)$ on host galaxy
extinction and SFH (Fig.~\ref{fig:highz}c). $z_{g',0.9}(\mglim)$ of the
estimates HB06-Gal and MDP07-Gal are consistent but they are slightly lower than the estimate
MDP07-no at $\mglim\gsim26$~mag and the estimate MK11-Gal at
$\mglim\lsim25$~mag. And the estimate HB06-LMC has slightly
higher $z_{g',0.9}(\mglim)$ at $\mglim\lsim25$~mag than the estimate
HB06-Gal, while $z_{g',0.9}(\mglim)$ of the estimate
HB06-LMC at $\mglim\gsim27$~mag and the estimate HB06-SMC are 
consistent with that of the estimate HB06-Gal. These are because $\zlim(\mlim)$
depends mainly on the host galaxy extinction at shallow $\mlim$ and the
IGM absorption at deep $\mlim$. The reason
why $z_{x,0.9}(\mlim)$ of the estimate MDP07-Gal is higher than that of
the estimate HB06-Gal at $\mlim\gsim31$~mag is because we assume the
constant SFR at high redshift in the estimate MDP07-Gal which is higher
than the SFR in \cite{hop06} at $z\gsim5$.

\subsubsection{Requirements on survey strategies}
\label{sec:realobs}

Although the above estimates refer only to the peak apparent magnitude,
the cadence of observations is also an important ingredient to identify
transients. Indeed, it is difficult to identify and interpret an event only
with one-epoch brightening. Additionally, the above estimates do not
take into account
an elongation of the duration at high redshift. Hence, we estimate the number of detection
of shock breakout $N_{i,x}(\mlim,\Omega_{\rm obs})$ for an
observation sampling $i$ with bandpass $x$, $\mlim$, and 
a field of view $\Omega_{\rm obs}$, which can be obtained by an integration of a ``control time'' 
$\Gamma_{i,x}(\Mms,A_{\rm V},z,\mlim)$\footnote{$\Gamma_{i,x}(\Mms,A_{\rm V},z,\mlim)$ 
is the number of
detection of shock breakout with $\Mms$, $A_{\rm V}$, and $z$ per unit SN rate
in the observer frame and thus having a dimension of number/(number/time),
\ie time.} as follows:
\begin{eqnarray}
 N_{i,x}(\mlim,\Omega_{\rm obs}) = \iiint \Gamma_{i,x}(\Mms,A_{\rm
  V},z,\mlim) \nonumber \\
\phi(\Mms)  {\Omega_{\rm obs}}
 {dV(z)\over{d\Omega dz}} {\eta(z)\over{1+z}} \chi(A_{\rm V}) d\Mms dz dA_{\rm V},
\end{eqnarray}
which is similar to Eq.~(\ref{eq:nx}) but integrates with 
$\Gamma_{i,x}(\Mms,A_{\rm V},z,\mlim)$ instead of 
$f_x[m_{{\rm peak},x}(\Mms,A_{\rm V},z),\mlim]$.
We define the detection with two criteria: (1) the event
is detected at $\geq N_{\rm detect}$ samplings with $3\sigma$ and (2) at
least one of samplings is taken from $\tobs=-0.2$~days to
$\tobs=0.4$~days.
$\Gamma_{i,x}(\Mms,A_{\rm V},z,\mlim)$ is determined by simulating the
detection of shock breakout. The expected number of detection and
reachable redshift are slightly smaller and lower than the previous
estimates in which all events with $m_{{\rm peak},x}\leq\mlim$
are counted because we require the detectable SN to be brighter than $\mlim$
over $\geq N_{\rm detect}$ samplings.

We attempt strategies A-N which divide a 6-, 12-, or 18-hour
$g'$-band observation with respective ways, assuming the cosmic SFH in
\cite{hop06} and our Galaxy extinction law \citep{pei92}. Here, we treat
the number of nights $N_{\rm night}$, observation sampling per night 
$n_{\rm obs}$, and exposure time for each sampling $t_{\rm exp}$ as
parameters, assume $N_{\rm detect}=3$, and adopt Subaru/Suprime-Cam to
estimate $\mlim$ for $3\sigma$ detection (\citealt{miy02}).
Each observation sampling is uniformly distributed to the whole night
($10$~hr for each night) and an overhead is neglected. 
The detail of strategies and the results, the expected number
per square degree
and the redshift below which $90\%$ of events occur, are summarized in
Table~\ref{tab:num}. 

While large $N_{\rm night}$
enhances the number of detection, large $n_{\rm obs}$ and long 
$t_{\rm exp}$ enhance the reachable redshift. For
the same $N_{\rm night}$, a strategy with $n_{\rm obs}=5$, in which a
field is observed every 2 hours, is the most efficient in
number and reachable redshift. 
The remarkably small number of
detection of strategy L ($n_{\rm obs}=1$) demonstrates that the multiple
photometric observation in a night is essential to detect shock
breakout because shock breakout brightens and declines within
$\sim1$~day. Comparing the strategies A, M, and N, an additional one
night observation increase the number of detection by
$2.3~{\rm degree^{-2}}$, which is larger than the number of detection of
the one night observation ($1.6~{\rm degree^{-2}}$, strategy A). This is because the observation
over $\geq2$ continuous nights can detect an event, taking place at the
end of the first night, at the subsequent nights. Thus, the
strategy with $N_{\rm night}\geq2$ is more favorable than the strategy
with $N_{\rm night}=1$. 

As expected in \S~\ref{sec:number}, a wider and shallower survey detects larger number of
shock breakout at lower redshift. However, the fraction of overhead
should be high for extremely short $t_{\rm exp}$ in reality. Thus, we
estimate the number of observable SNe taking into account
an overhead. Here, we expediently assume an overhead as $10\%$ of total
observation for $t_{\rm exp}\geq5$~min or $30$~sec for
$t_{\rm exp}<5$~min, and reduce the number of observable SNe in proportion to
the fraction of overhead. As a result, the number of observable SNe peaks at 
$t_{\rm exp}=5$~min for this specific overhead.

\subsubsection{Identification of shock breakout}
\label{sec:ID}

\begin{figure}[t]
\epsscale{1.}
\plotone{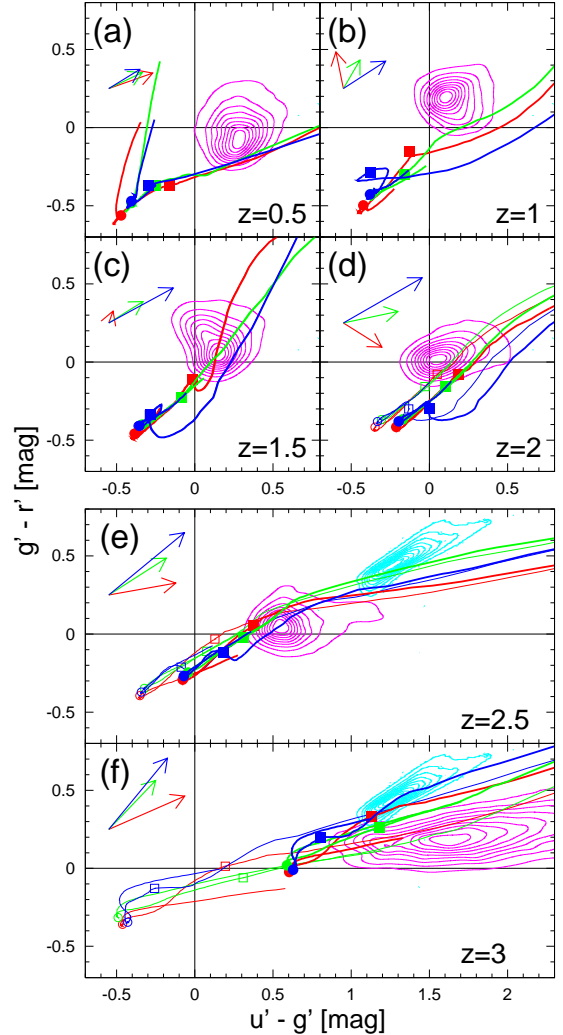} 
\caption{Color-color ($u'-g'$ vs. $g'-r'$) diagrams for the models with
 $\Mms=15\Msun$, $Z=0.02$, $E_{51}=1$ ({\it red}), $\Mms=25\Msun$,
 $Z=0.02$, $E_{51}=1$ ({\it green}), and $\Mms=25\Msun$,
 $Z=0.02$, $E_{51}=10$ ({\it blue}), at 
 (a) $z=0.5$, (b) $z=1$, (c) $z=1.5$, (d)
 $z=2$, (e) $z=2.5$, and (f) $z=3$, without the IGM absorption ({\it
 thin line}) and with the IGM absorption ({\it thick
 line}). The points represent the colors at $t=0$ ({\it circle}) and at 
 $t=2$~days in the observer frame ({\it square}). The contours show the
 distributions of stars ({\it cyan}, SDSS, \citealt{aba09}) and QSOs
 ({\it magenta}, \citealt{sch02,sch03,sch05,sch07,sch10}). The arrows show
 extinction vectors of host galaxies with $A_{\rm V}=0.25$~mag, assuming
 that the extinction curve is the same as our Galaxy ({\it red}), LMC
 ({\it green}), or SMC ({\it blue}) \citep{pei92}.
}
\label{fig:2colorugr}
\end{figure}

\begin{figure*}[t]
\epsscale{.8}
\plotone{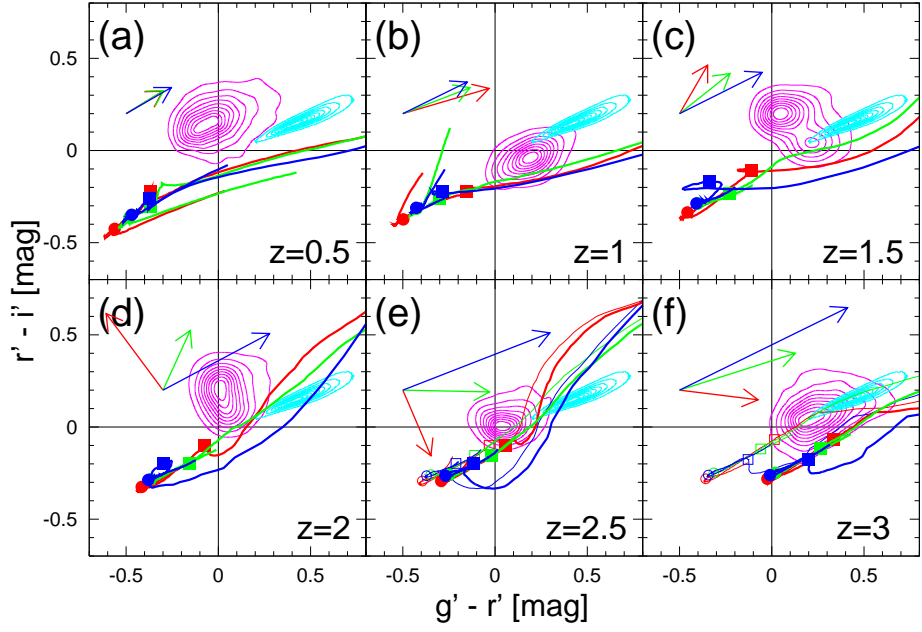} 
\caption{Color-color ($g'-r'$ vs. $r'-i'$) diagrams for the models at
 various redshifts. The panels, contours, points, lines, and arrows
 are the same as in Figure~\ref{fig:2colorugr} but $A_{\rm V}=0.5$~mag
 is assumed for the extinction vectors of host galaxies. 
}
\label{fig:2colorgri}
\end{figure*}

\begin{figure*}[t]
\epsscale{.8}
\plotone{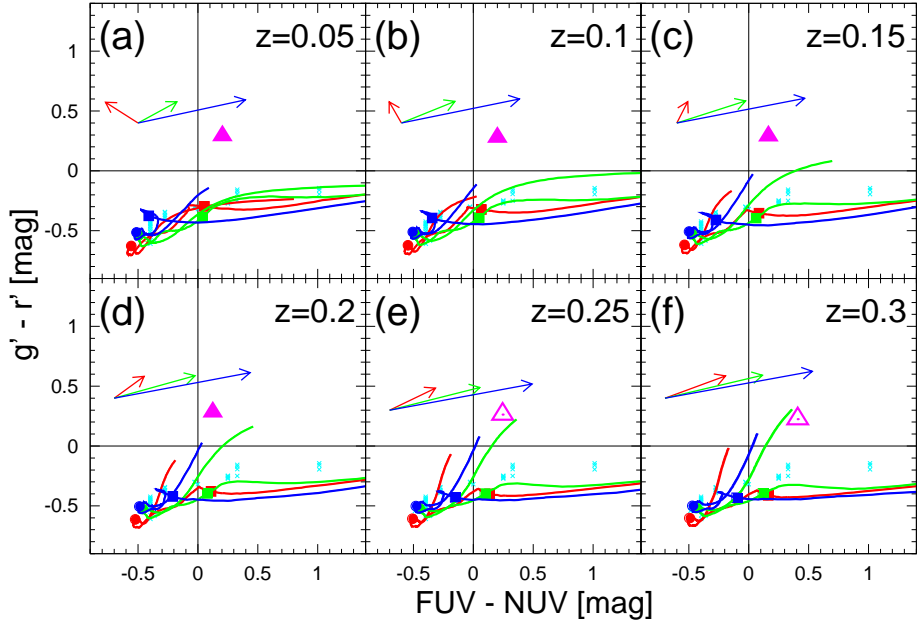} 
\caption{Color-color (FUV$-$NUV vs. $g'-r'$) diagrams of the models at
 (a) $z=0.05$, (b) $z=0.1$, (c) $z=0.15$, (d)
 $z=0.2$, (e) $z=0.25$, and (f) $z=0.3$. The points without lines represent
 colors of stars ({\it cyan}, Bruzual-Persson-Gunn-Stryker atlas,
 \citealt{str79,gun83}) and QSOs ({\it magenta}, \citealt{van01}). The
 colors of QSOs are represented by {\it open triangles} when the
 rest-frame Ly $\alpha$ wavelength is redshifted into FUV band. The
 color of lines, points on lines, and arrows are the same as in
 Figure~\ref{fig:2colorugr} but $A_{\rm V}=0.5$~mag is assumed for the
 extinction vectors of host galaxies. 
}
\label{fig:2colorUVgr}
\end{figure*}

\begin{figure*}[t]
\epsscale{.8}
\plotone{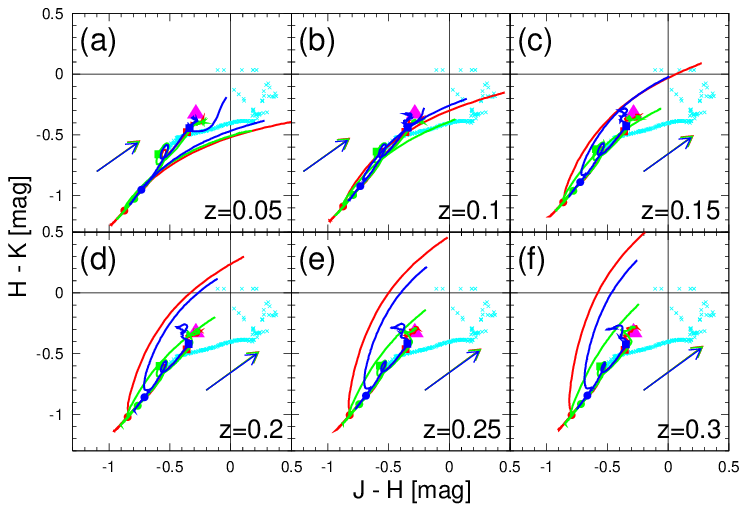} 
\caption{Color-color ($J-H$ vs. $H-K$) diagrams of the models at
 various redshifts. The panels, color of lines, points, and arrows are
 the same as in Figure~\ref{fig:2colorUVgr} but $A_{\rm V}=3$~mag
 is assumed for the extinction vectors of host galaxies. 
}
\label{fig:2colorJHK}
\end{figure*}

Many transients other than shock
breakout, \eg variable stars, SNe, quasars
(QSOs), and GRBs, will be found as variable objects in a
photometric observation. Shock breakout can be reliably discriminated 
from other kinds of variable objects holistically referring to
observable quantities: time scale, LC shape, color, and position, and
other observations. 

{\bf Time scale and LC shape}: Shock breakout has a nonrecurrent
brightening, a time scale of several sec $-$ several days, and featureless LC.
On the other hand, (1) SNe powered by radioactive decays, the
plateau of SNe II-P or Type IIn SNe, or linear decay of Type II-L SNe
have time scales of several ten days $-$ several hundred days, (2) GRB
optical flash has a time scale of several ten sec and frequently has a
jagged multiple peaks (\eg \citealt{woz09}), and (3) the flare of
low-mass star, \eg a M dwarf star, has a time scale of several minutes $-$ several hours and
recurrent brightening (\eg \citealt{haw03}). Therefore, the time scale
and LC shape of a transient can be employed for excluding these objects.

{\bf Position, archival image, and other observations}: If a survey is
performed at fields with plenty past observations,
checking past variabilities at the position
can effectively rule out possibilities of long time-scale variables such
as QSO or a variable star. 
If the event occurs in the outskirts of the host galaxy or is not
detected in X-ray, the variable object is likely to be shock
breakout.\footnote{Some low-luminosity AGN at high redshift can not
be detected even in deep X-ray data \citep{sar06,coh06agn,mor08}.}
And the deep UV, optical, and infrared imaging data are 
also useful for excluding the possibility of stars. 
Furthermore, if the field is included in the
field of view of $\gamma$-ray telescope (\eg {\it Swift}/BAT,
\citealt{geh04}), an alert of GRB can be
used for ruling out a possibility of GRB. Even if the GRB prompt
emission cannot be observed (\ie the GRB orphan afterglows), a radio
follow-up observation can constrain the presence of relativistic jets
(\eg \citealt{sod08,sod10}). 

{\bf Color}: The conclusive identification of shock breakout is given
by the color of a variable.
Figures~\ref{fig:2colorugr}a-\ref{fig:2colorugr}f ($u'-g'$ vs. $g'-r'$),
\ref{fig:2colorgri}a-\ref{fig:2colorgri}f ($g'-r'$ vs. $r'-i'$),
\ref{fig:2colorUVgr}a-\ref{fig:2colorUVgr}f (FUV$-$NUV vs. $g'-r'$), and
\ref{fig:2colorJHK}a-\ref{fig:2colorJHK}f ($J-H$ vs. $H-K$) show redshift-dependent
color-color diagrams for the less-massive $15\Msun$ and $E_{51}=1$
model, massive $25\Msun$ and $E_{51}=1$ model, and energetic $25\Msun$
and $E_{51}=10$ model with taking into account the IGM absorption
\citep{mad95} and compare them with color-color distributions of
stars\footnote{Photometrically-identified stars with photometric errors
$<0.01$~mag are extracted from the Seventh Data Release (DR7;
\citealt{aba09}) of the Sloan Digital Sky Survey (SDSS;
\citealt{yor00}). The number of stars is over $4\times10^6$. We note
that the sample could be biased toward bright and thus blue stars
because the stars with small errors are selected.} and QSOs
(\citealt{sch02,sch03,sch05,sch07,sch10}) for
Figures~\ref{fig:2colorugr}a-\ref{fig:2colorugr}f and
\ref{fig:2colorgri}a-\ref{fig:2colorgri}f and with colors derived from
typical spectra (stars: Bruzual-Persson-Gunn-Stryker
atlas,
\citealt{str79,gun83}\footnote{\url{http://www.stsci.edu/hst/observatory/cdbs/astronomical\_catalogs.html}}
and QSOs: \citealt{van01}) for Figures~\ref{fig:2colorUVgr}a-\ref{fig:2colorUVgr}f and 
\ref{fig:2colorJHK}a-\ref{fig:2colorJHK}f. Here, these figures show
extinction-uncorrected colors of stars and extinction-corrected colors of QSOs
at $z\pm\Delta z$ [where $\Delta z=0.1(1+z)$, being an accuracy of
a photometric redshift]. Ordinary SNe
are not shown because any type of SNe are too red to appear in these figures (\eg
\citealt{nug02}). The colors of shock breakout are FUV$-$NUV$\sim-0.5$, $u'-g'\lsim-0.3$,
$g'-r'\lsim-0.3$, $r'-i'\lsim-0.3$, $J-H\lsim-0.6$, and $H-K\lsim-0.8$
at $t=0$ and FUV$-$NUV$\sim0.1$, $u'-g'\lsim0.1$,
$g'-r'\lsim-0.1$, $r'-i'\lsim-0.1$, $J-H\lsim-0.3$, and $H-K\lsim-0.4$
at $\tobs=2$~days (see also Figs.~\ref{fig:gr}a-\ref{fig:gr}m), except for
$u'-g'$ at $z\geq2.5$ or $g'-r'$ at $z\geq3$. 
These exceptions stem from the fact that the light in
observed $u'$ and $g'$ bands are heavily absorbed by the IGM at
$z\geq2.5$ and $z\geq3$, respectively. The color of
shock breakout is much bluer than the majorities of stars and QSOs but
we note that the color at $t=0$ is similar to O/B stars and that the NIR
colors of less-massive and energetic models at $\tobs=2$~days and
massive model at $\tobs=10$~days are
similar to QSOs. 

GRB orphan afterglow and M dwarf flare could have a similar time scale,
featureless LC, and nonrecurrent brightening in a limited-time
photometric observation. However, the blue color is a precise
identifier of shock breakout.
According to a standard model for GRB afterglow 
\citep[\eg][]{sar98}, GRB afterglow are red at a frequency
range above frequencies corresponding to minimum or cooling Lorentz
factor. Adopting reasonable parameters for GRBs (\eg \citealt{pan02}),
the minimum or cooling frequency at $\gsim0.1$~days after the prompt
burst is $\nu\lsim10^{14}$~Hz and thus the red color is realized at
$\lambda\lsim3\times10^4$~\AA. Although the minimum and cooling
frequencies are higher for earlier epochs, the GRB orphan afterglow will
typically peak at $>0.1$~days (\eg \citealt{tot02}). Therefore, the GRB
orphan afterglow is likely to have a red color in optical bands. Also 
the color of low-mass star flare is typically red (\eg
\citealt{kow09}). 

\subsubsection{Constraints on supernova properties from shock breakout}
\label{sec:SNprop}

\begin{figure*}[t]
\epsscale{.95}
\plotone{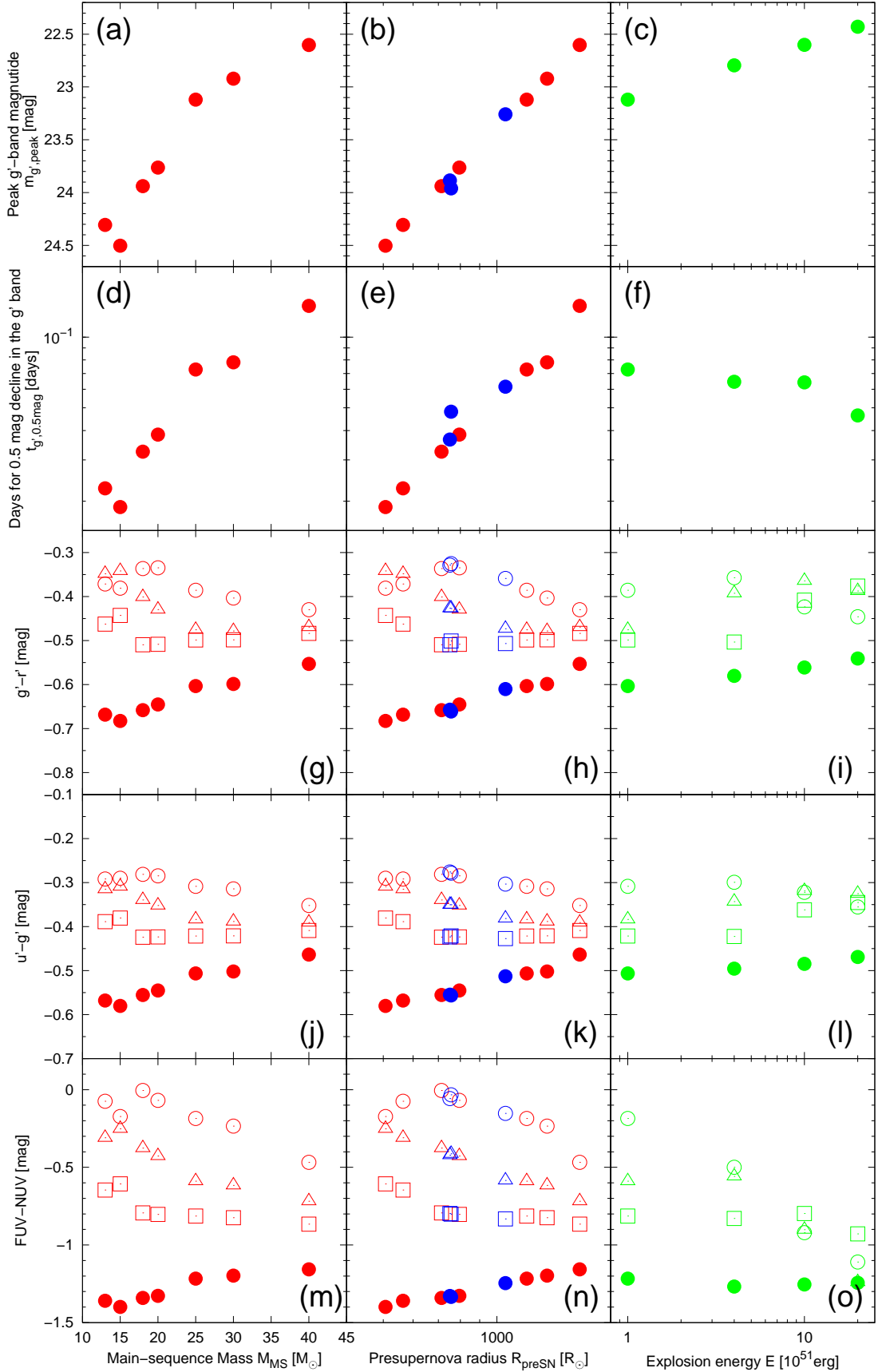}
\caption{Dependencies of observational quantities on the model properties,
 $\Mms$, $\Rs$, and $E$, at $z=0.2$. The color of symbols represents
 $Z=0.02$ models ({\it red}), $\Mms=20\Msun$ models with
 $Z=0.001,~0.004,~{\rm and}~0.05$ ({\it blue}), and $\Mms=25\Msun$
 models with $E_{51}=1,~4,~10,~{\rm and}~20$ ({\it green}). (g-o) The
 colors of the models at $\tbobs=0$ ({\it filled
 circles}), $0.5$~days ({\it open square}),
 $1$~day ({\it open triangles}), and $2$~days ({\it open circles}).
}
\label{fig:modelsz0.2}
\end{figure*}

\begin{figure*}[t]
\plotone{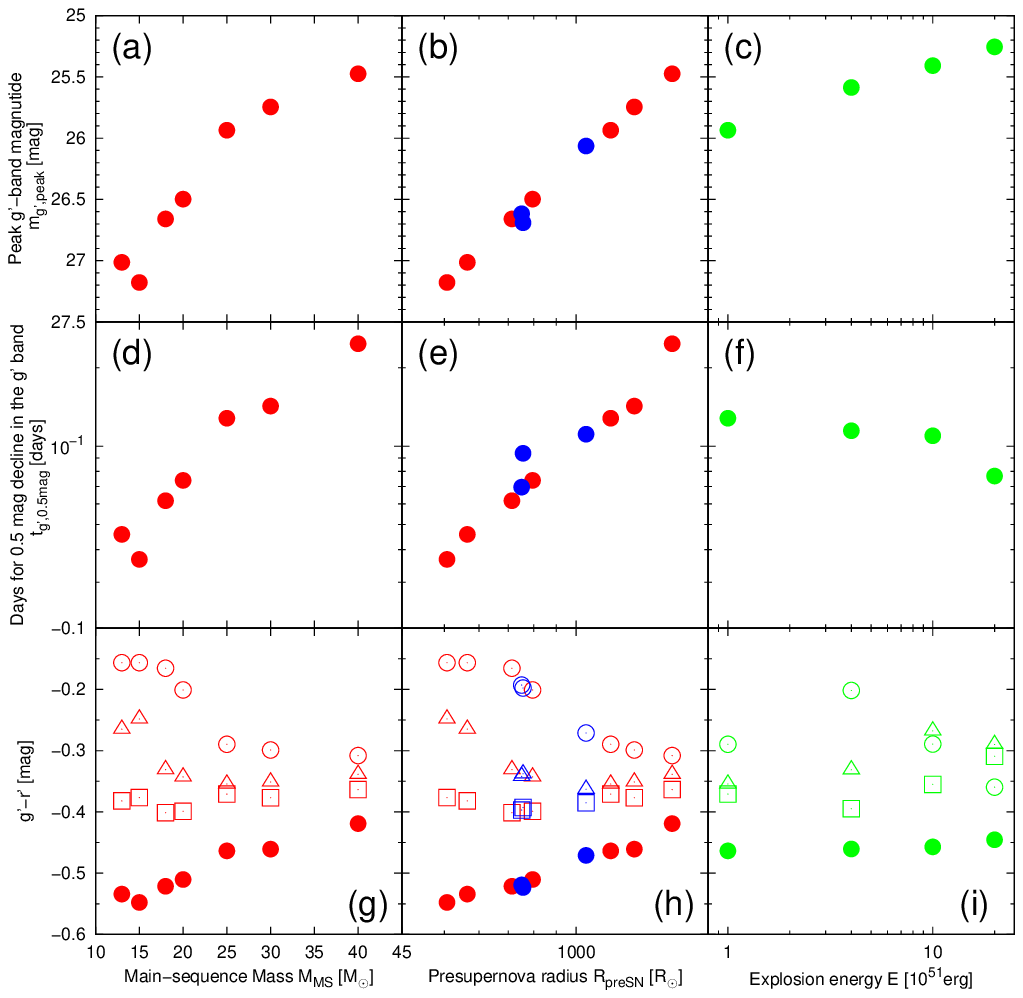}
\caption{Same as Figs.~\ref{fig:modelsz0.2}, but for $z=1$.
}
\label{fig:models}
\end{figure*}

\begin{figure*}[t]
\plotone{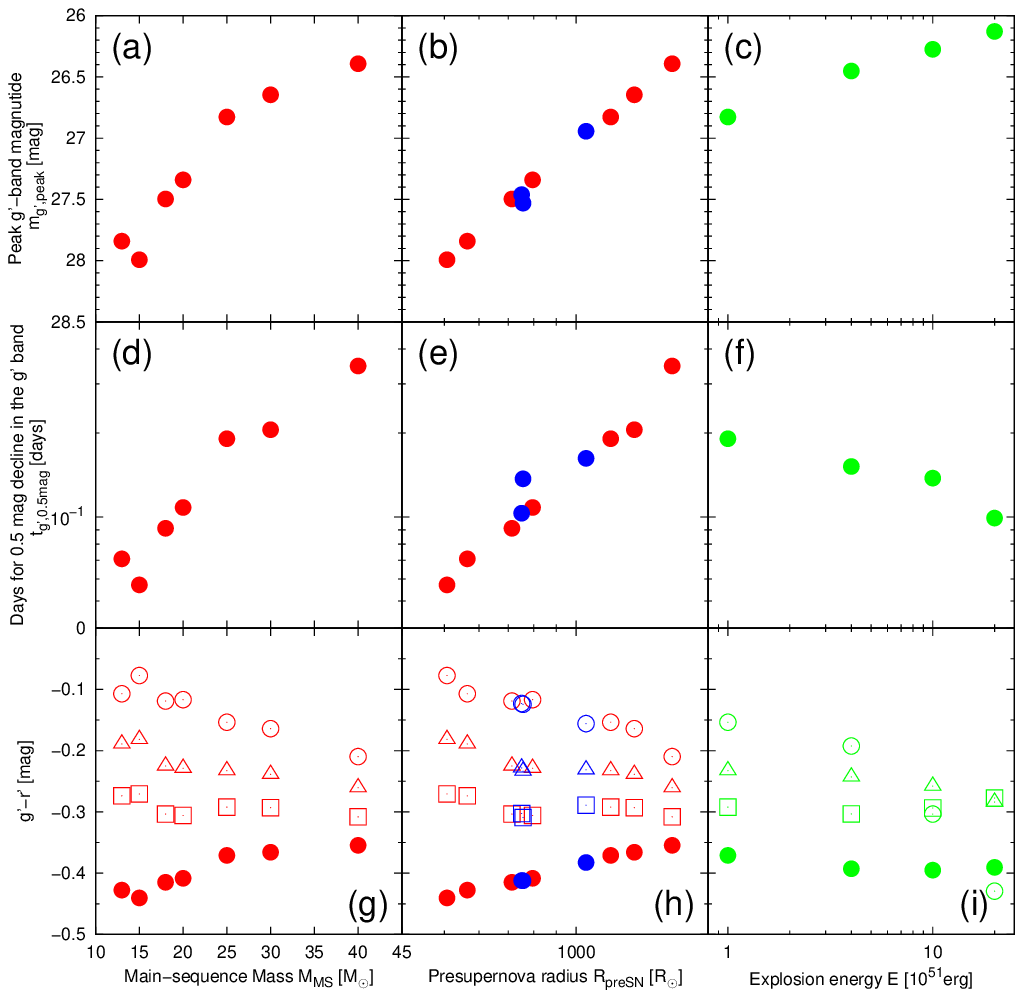}
\caption{Same as Figs.~\ref{fig:modelsz0.2}, but for $z=2$.
}
\label{fig:modelsz2}
\end{figure*}

The SN properties can be constrained from the observations of shock
breakout. Here, we focus on three observational quantities: peak
magnitude, decline rate, and color evolution. Their dependencies on the
model properties, $\Mms$, $\Rs$, and $E$, at $z=0.2$, $z=1$, and $z=2$ are
summarized in Tables~\ref{tab:obsz0.2}, \ref{tab:obsz1}, \ref{tab:obsz2}
and shown in Figures~\ref{fig:modelsz0.2}a-\ref{fig:modelsz0.2}o,
\ref{fig:models}a-\ref{fig:models}i, and
\ref{fig:modelsz2}a-\ref{fig:modelsz2}i, respectively.
Here, we assume that the redshift of shock breakout is determined 
photometrically or spectroscopically.

Figures~\ref{fig:modelsz0.2}a-\ref{fig:modelsz0.2}c,
\ref{fig:models}a-\ref{fig:models}c, and
\ref{fig:modelsz2}a-\ref{fig:modelsz2}c show the apparent peak $g'$-band
magnitude $m_{g',{\rm peak}}$. $m_{g',{\rm peak}}$ vary over $\sim1.5-2$~mag
depending on $\Rs$ and thus $\Mms$, although $\Lp$ are similar for the
models with $E_{51}=1$ (Fig.~\ref{fig:SB}a). On
the other hand, although $\Lp$ varies by an order of magnitude
depending on $E$ (Fig.~\ref{fig:SB}b), the $m_{g',{\rm peak}}$ range of
the models with different $E$ is only
$\sim0.7$~mag. The different behavior between bolometric and monochromatic
luminosities stems from the different photospheric temperatures leading
to different $T_{\rm c}$. The models with larger
$\Rs$ have lower temperature and thus redder SEDs and slightly brighter at
$\lambda\gsim300$\AA\ (Fig.~\ref{fig:color}a), while the models with higher $E$
have higher temperature and thus bluer SEDs but the luminosities are
almost similar at $\lambda\gsim800$\AA\ 
(Fig.~\ref{fig:color}b). We note that the $g'$-band luminosities peak at
$\tobs\sim0.001-0.05$~days because of the
temperature evolution (see also Figs.~\ref{fig:gLC}a-\ref{fig:gLC}m).

Figures~\ref{fig:modelsz0.2}d-\ref{fig:modelsz0.2}f,
\ref{fig:models}d-\ref{fig:models}f, and
\ref{fig:modelsz2}d-\ref{fig:modelsz2}f show days after the $g'$-band
peak until $g'$-band magnitude declines by $0.5$ magnitude $t_{g',{\rm 0.5mag}}$.
The dependencies of the decline rate in $g'$ band on $\Rs$ ($\Mms$) and
$E$ are almost similar to those in the bolometric LC. This
is because the decline rate depends on energetics of SNe but not on the
temperature when the SED peaks at shorter $\lambda$ than $g'$ band. Since the
dependencies of $m_{g',{\rm peak}}$ and $t_{g',{\rm 0.5mag}}$ on $\Rs$
($\Mms$) and $E$ are different, in principal they could determine $\Rs$
($\Mms$) and $E$ independently.

However, apparent brightness is dimmed by extinction and
thus having large uncertainties. Hence, we introduce the color
evolution to resolve the uncertainties. Although the absolute
color is also strongly reddened by extinction, the color evolution
does not suffer from extinction unless extinction 
changes with time. Figures~\ref{fig:modelsz0.2}g-\ref{fig:modelsz0.2}o,
\ref{fig:models}g-\ref{fig:models}i, and
\ref{fig:modelsz2}g-\ref{fig:modelsz2}i show the color evolutions, FUV-NUV,
$u'-g'$, and $g'-r'$ for $z=0.2$ and $g'-r'$ for $z=1$ and $z=2$, respectively
(see also Figs.~\ref{fig:gr}a-\ref{fig:gr}m). These figures show
the bluest color, that is realized at $\tbobs=0$ 
corresponding to $\tobs\sim0.03-0.2$~days,
and the colors at $\tbobs=0.5$, $1$, and $2$ days, where $color$ is FUV-NUV,
$u'-g'$, or $g'-r'$. The colors of some models evolve
toward red at first and then get back to blue again, which is shown as
a loop structure in Figures~\ref{fig:2colorugr}a-\ref{fig:2colorugr}f,
\ref{fig:2colorgri}a-\ref{fig:2colorgri}f,
\ref{fig:2colorUVgr}a-\ref{fig:2colorUVgr}f, and
\ref{fig:2colorJHK}a-\ref{fig:2colorJHK}f.

The model with smaller $\Rs$ has a bluer color at the peak
because of the higher temperature and its color evolution is more rapid. 
The color evolution within 2 days are $\Delta(g'-r')\gsim0.3$~mag for the
model with $\Rs<10^3\Rsun$ and $\Delta(g'-r')\lsim0.25$~mag for the
model with $\Rs>10^3\Rsun$. For example, the $13\Msun$ model has
a redder color than the $30\Msun$ model at $\tbobs=1$~day for
$z=0.2$ and at $\tgrobs=2$~days for $z=1$ and $z=2$. 
On the other hand, the bluest colors of models with different $E$
are similar but the models with higher $E$ have more rapid and
smaller variations. The color evolutions of the models with
different $Z$ are similar to those of the
$Z=0.02$ models with the same $\Rs$ and thus it could cause an
uncertainty for constraining $\Mms$ when the metallicity of host galaxy is
unknown. In conclusion, the color evolutions can classify shock
breakout, at least, to the following three groups: explosions with
$\Rs<10^3\Rsun$ and $\Rs>10^3\Rsun$ ($\Mms\leq20\Msun$ and
$\Mms\geq25\Msun$ for the $Z=0.02$ models) and an energetic explosion
with $E_{51}\geq10$. 

\section{CONCLUSIONS \& DISCUSSION}
\label{sec:discuss} 

Shock breakout is the brightest event in the SN with a shockwave and could
consummate the detection of CCSNe in the high-$z$ universe. We
present multicolor LCs of shock breakout in SNe~II-P with various
$\Mms$, $Z$, and $E$ based on realistic stellar models. Using our
theoretical models, we 
investigate the dependencies of shock breakout properties on the
progenitors and explosion energies and present
thorough prospects for future surveys of shock breakout. It is
essential for identifying and interpreting shock breakout to
observe a field more than once in a night 
in multiple blue optical bands, preferably over $\geq2$ continuous
nights. And, adopting standard cosmic SFH, IMF,
extinction distribution of host galaxies, the $g'$-band observable SN rate for
$\mglim=27.5$~mag is $3.3$~SNe~degree$^{-2}$~day$^{-1}$ and the half of them
locates at $z\geq1.2$.

We calculate 13 SN models with $\Mms=13-40\Msun$, $Z=0.001-0.05$, and
$E_{51}=1-20$ (Tab.~\ref{tab:SB}). The model with larger $\Rs$, thus
typically larger $\Mms$, has
lower $\Tp$, longer $\thalf$, and higher $\Erad$, while the model with
higher $E$ has higher $\Tp$, $\Erad$, and $\Lp$. The metallicity affects
shock breakout mainly through altering the stellar structure. The variations
of $\Tp$, $\thalf$, $\Erad$, and $\Lp$ among the adopted models are
$\sim(2-5)\times10^5$~K, $\sim0.01-0.6$~days,
$\sim(0.7-5)\times10^{48}$~erg, and $\sim(0.5-7)\times10^{44}$~\ergs,
respectively. The dependencies of numerical results are similar to
those of the semi-analytic solutions \citep{mat99}. The semi-analytic
solutions are nearly consistent with the numerical results if
$\Tmm$, $\Emm$, $\tmm$, and $\Lmm$ are reduced by factors of 1.5, 3, 2,
and 1.5, respectively. 

According to the models, we predict the observational
quantities of high-$z$ shock breakout.
Since shock breakout has a blue SED peaked at $\sim100$~\AA, 
brightness of shock breakout at a fixed observed bandpass is less
dimmed compared to the geometrical dilution, \ie shock breakout has
a large negative $K$-correction. This makes it possible to detect 
high-$z$ shock breakout, \eg up to $z\sim3.5$ in $g'$ band with
8m-class telescopes if there is no extinction in
the host galaxy. Although shock breakout strongly suffers
from the extinction and IGM absorption, it can be detected up to
$z\sim2$ if the host galaxy has the color excess $\Ebvh=0.1$~mag and our Galaxy
extinction law.

Convolving the Salpeter's IMF, cosmic SFH, host galaxy extinction, and
IGM absorption, we estimate the observable SN rate as a function of
bandpass and limiting magnitude. As a result, considering the operative
telescopes/instruments, the $g'$ band observation
is currently the most effective in the number of detection. Adopting
cosmic SFH by \cite{hop06} and our Galaxy extinction law, the observable
SN rate is $3.3$ SNe~degree$^{-2}$~day$^{-1}$ for $\mglim=27.5$~mag.
Even taking into account uncertainties on host galaxy extinction and
SFH, the observable SN rate is $\geq0.93$ SNe~degree$^{-2}$~day$^{-1}$. 

We also present the redshift
distribution of observable SNe. For currently-available
$\mglim=27.5$~mag, $50\%$ of observable SNe take place
at $z\geq1.2$. Furthermore, for $\mglim=30$~mag, $\sim10\%$ of observable
SNe locate at $z\geq3$. 
Since the reachable redshift increases dramatically if $\mlim\gsim26-30$~mag is
feasible, the next-generation telescopes/instruments will considerably
enhance the reachable redshift.
The reachable redshift is almost independent of
the uncertainties involved in host galaxy extinction and SFH. Therefore, the
shock breakout is the most appropriate phenomenon to aim at the
detection of high-$z$ CCSNe. The first detection of normal CCSNe at
$z>1$ can certainly be achieved by the observation of shock breakout in SNe~II-P.
The direct observation of normal CCSNe at $z>1$ will shed light
on their nature and cosmic evolution histories, most of
which are currently derived from galaxy studies that might be biased by
brightness of galaxies.

Future/ongoing wide and/or deep surveys, \eg Palomar Transient
Factory (PTF, \citealt{law09,rau09}), Lick Observatory Supernova Search
(LOSS, \citealt{lea10}), Catalina Real-Time Transient Survey (CRTS,
\citealt{dra09}), Kiso/Kiso Wide Field Camera
(KWFC),\footnote{\url{http://www.ioa.s.u-tokyo.ac.jp/kisohp/top\_e.html}}
Skymapper,\footnote{\url{http://www.mso.anu.edu.au/skymapper/}} Dark Energy
Survey (DES, \citealt{ber09}), Panoramic Survey Telescope and Rapid
Response System (Pan-STARRS, \citealt{kai02,kai04}), Subaru/Hyper
Suprime-Cam (HSC, \citealt{miy06}), and Large Synoptic Survey Telescope
(LSST, \citealt{ive08}), will find a large number of shock breakout. 
We simulate realistic survey strategies and show that a wider and
shallower survey leads to a higher observable SN rate with a given survey
power but misses the higher-$z$ events. And the observable SN rate for
short integration is suppressed by the overhead. Although the survey
parameters should be customized to observation purposes and
telescope/instrument, we conclude that the most essential observation is
the multicolor photometry with short intervals less than 1 day and that
the observation over $\geq2$ continuous nights is favorable.

We also establish the ways to identify shock breakout and to
constrain the SN properties from the observations of shock breakout.
The LC, color, position, and past or
wide frequency-coverage observations excellently distinguish shock
breakout from the variable stars, SNe, GRBs, and QSOs. In
particular, the blue color of shock breakout is the most important
information to identify shock breakout. Shock breakout in
an SN~II-P with larger $\Rs$ evolves more slowly and has a more luminous peak in
optical bandpass. The two observational quantities, time scale and color
evolution, can reasonably determine the SN properties, $\Rs$ ($\Mms$) and $E$, being
independent of the host galaxy extinction. When numerous shock
breakout is detected, the IMF in the high-$z$ universe will be
constrained by shock breakout. Furthermore, if the SN properties can be
determined only from the time scale and color variation with time, the other two
observational quantities, peak magnitude and absolute color, can determine the
host galaxy extinction.  Combining the observations
of the host galaxies, the relation between the host galaxy and its
stellar contents might be constrained.

Shock breakout enables an untargeted CCSN survey at unprecedentedly high
redshift, which can provide large uniform CCSN samples with
the comparable redshift range with GRBs. This would allow us to clarify the relation between
GRBs and star formation and thus the GRB progenitors. GRBs are hosted
in blue, faint, and/or low-$Z$ galaxies \citep[][]{lef03,kew07,lev10b}
and thus have been suggested to require low-$Z$ progenitors. On the other
hand, \cite{koc10} and \cite{man10b} recently suggest that the GRB hosts
follow a tight correlation among stellar mass, metallicity, and SFR of
field galaxies and that the characteristics of GRB hosts can be explained only by
the large SFR in low-$Z$ galaxies. If the latter is correct, the host
galaxies of CCSN should share the same properties as the GRB
hosts because the lifetimes of GRB and CCSN progenitors are
similar.\footnote{For example, the lifetimes of stars with
$\Mms\geq40\Msun$, a putative GRB progenitor, and $\Mms\geq10\Msun$,
a putative CCSN progenitor, are less than $5$~Myr and $20$~Myr,
respectively \citep{sch92}. In order to make a difference between GRB and
CCSN hosts, the star formation in the host galaxies
should cease $5-20$~Myr ago or start $\leq5$~Myr ago. Although it is
still under debate how short duration of star formation activity
is allowed \citep[\eg][]{mas99,mcq10}, such short time scales are
comparable with the age spreads of OB associations and the lifetimes of
giant molecular clouds (\citealt{mck07} and references therein). }
However, it is suggested that the GRB hosts are fainter and
more irregular at $z<1.2$ and lower-$Z$ at $z<0.3$ than the CCSN hosts
\citep{fru06,mod08}, although there are rooms for improvement, \eg the
statistics and uniformity. Hence, the large CCSN sample of the same
quality as GRBs, especially on survey method and redshift range, could
give unique information on the environment of GRBs. The comparison
between CCSN and GRB hosts at $z>1$ can provide an essential clue to
unpuzzle the issue. It will probe whether an SFH estimate with GRBs,
which is extendible to $z\gsim8$, are biased or not.

Aspherical shock breakout in a cocoon or a jet is suggested
for Type Ic SN~2006aj/GRB~060218 \citep[\eg][]{sod06,cam06,wax07,ghi07}
and Type Ib SN~2008D/XRF~080109 \citep[\eg][]{sod08,maz08}. The
asphericity stems from the compactness of the progenitor which leads to
relativistic outflow at the tip of shockwave.
Therefore, in order to precisely deal with
aspherical shock breakout, a multidimensional relativistic radiation
hydrodynamics calculation is required but has not been attained so far.
On the other hand, polarization observations demonstrate that
SNe~II-P we focused have a spherical structure at the plateau phase even with
an aspherical inner core \citep[\eg][]{leo06,wang08}. This is because the
progenitor of an SN~II-P has a thick H envelope diluting the asphericity. 
Therefore, assuming the spherical symmetry is reasonable for SNe~II-P
and the results are applicable for all SNe~II-P.

Recently \cite{sma09a} suggests that there is a maximum mass for the
progenitor that can be SNe~II-P
($\Mms\leq16.5\pm1.5\Msun$).\footnote{\cite{des10} also suggests that this constraint is
consistent with the nebular observations of SNe~II-P.} This suggestion 
cautions the existence of a massive SNe~II-P with $\Mms>20\Msun$.
However, the constraints are limited only to SNe~II-P having
occurred in nearby universe, in which individual stars can be
resolved. It is not investigated whether such a low maximum mass for SNe~II-P
exists in high-$z$ universe or at low-$Z$ environment. 
Also, the reason why the stars with $\Mms>20\Msun$ cannot be
SNe~II-P is under debate. Some studies suggested that such a massive
star forms a black hole directly (\eg \citealt{sma09b}) or explodes as
other kinds of SNe due to a strong mass loss and/or rotation (\eg
\citealt{sma09a,yoo10}, S. Ekstr\"{o}m et al. in prep.). However, both
scenarios are not conclusive because some
of stars with $\Mms\geq25\Msun$ explode as energetic Type Ic SNe with
$E_{51}\geq10$ (\eg \citealt{iwa98}), which is enough to explode a massive star
even with a thick H envelope, and because the mass loss from a massive star is 
difficult to be predicted theoretically (\eg
\citealt{vin08} for review). Furthermore, a fast rotating star, that can
explain the low maximum mass, has a larger presupernova radius at a
given mass than a non-rotating star if the H envelope remains
(C. Georgy, S. Ekstr\"{o}m and G. Meynet, private communication). This
could moderate the reduction of the observable SN rate.
The observational properties of shock breakout can constrain the properties of SNe~II-P
and their progenitors at high redshift. The redshift-dependent
number count and properties of shock breakout can judge when the maximum mass 
for SNe~II-P is established and could give a clue to the
origin of the maximum mass for SNe~II-P.

\acknowledgments

The authors appreciate help by Hideyuki Umeda,
A.W.A. Pauldrach, and Masakazu A.R. Kobayashi for providing a progenitor
model, an atomic data, and a semi-analytic model, respectively.
N.T. and T.M. thank Nobunari Kashikawa for
valuable discussion on IGM. N.T. also thank Masaomi Tanaka, Keiichi
Maeda, Robert Quimby, Hajime Susa, Cyril Georgy, Sylvia Ekstr\"{o}m, and
Georges Meynet for fruitful discussion on host galaxies, radiation
transfer, transients, star formation, and presupernova
structures. T.M. has been supported by the JSPS (Japan Society for the
Promotion of Science) Research Fellowship for Young Scientists.
The work of S.B., P.B., and E.S. in Russia is supported partly by the grant
RFBR 10-02-00249-a,
by Scientific School Foundation grants 2977.2008.2, 3884.2008.2,
by contract with Agency for Science and Innovation No.~02.740.11.0250,
contract with Rosnauka No.~02.740.11.5158,
SNSF grant  No.~IZ73Z0-128180/1 under the program SCOPES, and in Germany by
MPA guest program.
This research has been supported in part by World Premier
International Research Center Initiative, MEXT,
Japan, and by the Grant-in-Aid for Scientific Research of the JSPS
(18104003, 20540226, 21840055) and MEXT (19047004, 22012003).

\bibliographystyle{apj} 
\bibliography{ms}

\clearpage

\hoffset=-2cm
\textwidth=7.6in

\begin{deluxetable*}{ccc|cccccccc}[t]
 \tabletypesize{\tiny}
 \tablecaption{Observable quantities at $z=0.2$. \label{tab:obsz0.2}}
 \tablewidth{0pt}
 \tablehead{
   \colhead{$\Mms$}
 & \colhead{$Z$}
 & \colhead{$E$}
 & \colhead{$m_{g',{\rm peak}}$}
 & \colhead{$t_{g',{\rm 0.5mag}}$}
 & \multicolumn{2}{c}{FUV$-$NUV}
 & \multicolumn{2}{c}{$u'-g'$}
 & \multicolumn{2}{c}{$g'-r'$}\\
   \colhead{[$\Msun$]}
 & \colhead{}
 & \colhead{[$10^{51}$~erg]}
 & \colhead{[mag]}
 & \colhead{[$0.01$~days]}
 & \multicolumn{2}{c}{[mag]}
 & \multicolumn{2}{c}{[mag]}
 & \multicolumn{2}{c}{[mag]}\\
 & \colhead{}
 & \colhead{}
 & \colhead{}
 & \colhead{}
 & \colhead{$t^{\rm FUV-NUV}_{{\rm obs}}=0$}
 & \colhead{$t^{\rm FUV-NUV}_{{\rm obs}}=2$~days}
 & \colhead{$t^{u'-g'}_{{\rm obs}}=0$}
 & \colhead{$t^{u'-g'}_{{\rm obs}}=2$~days}
 & \colhead{$\tgrobs=0$}
 & \colhead{$\tgrobs=2$~days}
 }

\startdata
 13& 0.02  &  1 & 24.31 & 2.27 & -0.668 & -0.371 & -0.568 & -0.292 & -1.45 & -0.0750 \\
 15& 0.02  &  1 & 24.50 & 1.89 & -0.683 & -0.381 & -0.580 & -0.290 & -1.49 & -0.173 \\
 18& 0.02  &  1 & 23.94 & 3.25 & -0.658 & -0.337 & -0.556 & -0.281 & -1.41 & -0.00539 \\
 20& 0.02  &  1 & 23.76 & 3.85 & -0.645 & -0.335 & -0.545 & -0.285 & -1.38 & -0.0692\\
 25& 0.02  &  1 & 23.12 & 7.28 & -0.603 & -0.386 & -0.507 & -0.308 & -1.24 & -0.185 \\
 30& 0.02  &  1 & 22.92 & 7.81 & -0.599 & -0.403 & -0.502 & -0.314 & -1.23 & -0.235 \\
 40& 0.02  &  1 & 22.60 & 13.6 & -0.553 & -0.430 & -0.463 & -0.352 & -1.17 & -0.468 \\ 
 25& 0.02  &  4 & 22.80 & 6.46 & -0.580 & -0.357 & -0.495 & -0.299 & -1.30 & -0.499\\
 25& 0.02  & 10 & 22.60 & 6.42 & -0.561 & -0.424 & -0.485 & -0.322 & -1.31 & -0.921 \\
 25& 0.02  & 20 & 22.43 & 4.64 & -0.541 & -0.446 & -0.469 & -0.356 & -1.29 & -1.11\\
 20& 0.001 &  1 & 23.96 & 4.81 & -0.661 & -0.325 & -0.557 & -0.279 & -1.39 & -0.0328\\
 20& 0.004 &  1 & 23.89 & 3.66 & -0.657 & -0.329 & -0.555 & -0.276 & -1.39 & -0.0569 \\
 20& 0.05  &  1 & 23.26 & 6.15 & -0.610 & -0.359 & -0.513 & -0.303 & -1.28 & -0.152 
\enddata

\end{deluxetable*}

\begin{deluxetable*}{ccc|cccc}[t]
 \tabletypesize{\scriptsize}
 \tablecaption{Observable quantities at $z=1$. \label{tab:obsz1}}
 \tablewidth{0pt}
 \tablehead{
   \colhead{$\Mms$}
 & \colhead{$Z$}
 & \colhead{$E$}
 & \colhead{$m_{g',{\rm peak}}$}
 & \colhead{$t_{g',{\rm 0.5mag}}$}
 & \multicolumn{2}{c}{$g'-r'$}\\
   \colhead{[$\Msun$]}
 & \colhead{}
 & \colhead{[$10^{51}$~erg]}
 & \colhead{[mag]}
 & \colhead{[$0.01$~days]}
 & \multicolumn{2}{c}{[mag]}\\
 & \colhead{}
 & \colhead{}
 & \colhead{}
 & \colhead{}
 & \colhead{$\tgrobs=0$}
 & \colhead{$\tgrobs=2$~days}
 }
\startdata
 13& 0.02  &  1 & 27.01 & 4.58 & -0.534 & -0.156  \\
 15& 0.02  &  1 & 27.17 & 3.67 & -0.548 & -0.156  \\
 18& 0.02  &  1 & 26.66 & 6.18 & -0.521 & -0.165  \\
 20& 0.02  &  1 & 26.50 & 7.39 & -0.511 & -0.201  \\
 25& 0.02  &  1 & 25.94 & 12.8 & -0.464 & -0.290  \\
 30& 0.02  &  1 & 25.74 & 14.3 & -0.461 & -0.299  \\
 40& 0.02  &  1 & 25.48 & 24.7 & -0.419 & -0.308  \\ 
 25& 0.02  &  4 & 25.59 & 11.5 & -0.460 & -0.202  \\
 25& 0.02  & 10 & 25.41 & 11.0 & -0.457 & -0.290  \\
 25& 0.02  & 20 & 25.26 & 7.67 & -0.446 & -0.360  \\
 20& 0.001 &  1 & 26.69 & 9.39 & -0.523 & -0.198  \\
 20& 0.004 &  1 & 26.62 & 6.96 & -0.519 & -0.193  \\
 20& 0.05  &  1 & 26.06 & 11.1 & -0.471 & -0.271 
\enddata

\end{deluxetable*}

\begin{deluxetable*}{ccc|cccc}[t]
 \tabletypesize{\scriptsize}
 \tablecaption{Observable quantities at $z=2$. \label{tab:obsz2}}
 \tablewidth{0pt}
 \tablehead{
   \colhead{$\Mms$}
 & \colhead{$Z$}
 & \colhead{$E$}
 & \colhead{$m_{g',{\rm peak}}$}
 & \colhead{$t_{g',{\rm 0.5mag}}$}
 & \multicolumn{2}{c}{$g'-r'$}\\
   \colhead{[$\Msun$]}
 & \colhead{}
 & \colhead{[$10^{51}$~erg]}
 & \colhead{[mag]}
 & \colhead{[$0.01$~days]}
 & \multicolumn{2}{c}{[mag]}\\
 & \colhead{}
 & \colhead{}
 & \colhead{}
 & \colhead{}
 & \colhead{$\tgrobs=0$}
 & \colhead{$\tgrobs=2$~days}
 }
\startdata
 13& 0.02  &  1 & 27.84 & 7.08 & -0.428 & -0.107  \\
 15& 0.02  &  1 & 27.99 & 5.71 & -0.440 & -0.077  \\
 18& 0.02  &  1 & 27.50 & 9.11 & -0.415 & -0.119  \\
 20& 0.02  &  1 & 27.34 & 10.8 & -0.408 & -0.117  \\
 25& 0.02  &  1 & 26.83 & 19.1 & -0.371 & -0.154  \\
 30& 0.02  &  1 & 26.65 & 20.6 & -0.366 & -0.164  \\
 40& 0.02  &  1 & 26.39 & 34.8 & -0.355 & -0.210  \\
 25& 0.02  &  4 & 26.45 & 15.2 & -0.393 & -0.193  \\
 25& 0.02  & 10 & 26.28 & 13.8 & -0.395 & -0.304  \\
 25& 0.02  & 20 & 26.13 & 9.92 & -0.391 & -0.429  \\
 20& 0.001 &  1 & 27.53 & 13.7 & -0.413 & -0.125  \\
 20& 0.004 &  1 & 27.46 & 10.3 & -0.412 & -0.123  \\
 20& 0.05  &  1 & 26.94 & 16.2 & -0.383 & -0.156 
\enddata

\end{deluxetable*}

\end{document}